\documentclass{bmcart}

\usepackage{amsthm,amsmath}
\usepackage[utf8]{inputenc} 

\usepackage{graphicx}

\begin{document}

\begin{frontmatter}

\begin{fmbox}
\dochead{Research}


\title{Creation of two-dimensional coulomb crystals of ions in oblate Paul traps for quantum
  simulations}


\author[
   addressref={aff1},
   corref={aff1},
   email={bty4@georgetown.edu}   
]{\inits{BTY}\fnm{Bryce} \snm{Yoshimura}}
\author[
   addressref={aff2},
   email={marybeth.stork@gmail.com}
]{\inits{MS}\fnm{Marybeth} \snm{Stork}}
\author[
	 addressref={aff3},
	 email={danilodadic@gmail.com},
]{\inits{DD}\fnm{Danilo} \snm{Dadic}}
\author[
	 addressref={aff3},
	 email={wes@physics.ucla.edu},
]{\inits{WCC}\fnm{Wesley C.} \snm{Campbell}}
\author[
	 addressref={aff1},
	 email={freericks@physics.georgetown.edu}
]{\inits{JKF}\fnm{J. K.} \snm{Freericks}}


\address[id=aff1]{
  \orgname{Department of Physics, Georgetown University}, 
  \street{37th and O St. NW},                     %
  \postcode{20007},                                
  \city{Washington, DC},                              
  \cny{USA}                                    
}
\address[id=aff2]{%
  \orgname{Department of Physics and Astronomy, Washington University},
  \street{Campus Box 1105, One Brookings Dr.},
  \postcode{63130},
  \city{St. Louis, Missouri},
  \cny{USA}
}

\address[id=aff3]{ \orgname{University of California Los Angeles}, \street{475 Portola Plaza},
\postcode{90095}, \city{Los Angeles, CA}, \cny{USA} }


\end{fmbox}


\begin{abstractbox}

\begin{abstract} 
We develop the theory to describe the equilibrium ion positions and
phonon modes for a trapped ion
quantum simulator in an oblate Paul trap that creates two-dimensional
Coulomb crystals in a triangular lattice. By coupling the internal states of the ions to 
laser beams propagating along the symmetry axis,   we study the effective
Ising spin-spin interactions that are mediated via the axial phonons and are less sensitive
to ion micromotion.  We find that the axial mode frequencies
permit the programming of Ising interactions with inverse power law
spin-spin couplings that can be tuned from uniform to $r^{-3}$ with DC
voltages.  Such a trap could allow for interesting new geometrical
configurations for quantum simulations on moderately sized systems
including frustrated magnetism on triangular lattices or Aharonov-Bohm effects
on ion tunneling.  The trap also
incorporates periodic boundary conditions around loops which could be employed to examine time crystals.

\end{abstract}




\end{abstractbox}
%

\end{frontmatter}




\section{Introduction}

Using a digital computer to predict the ground state of complex
many-body quantum systems, such as frustrated magnets, becomes an
intractable problem when the number of spins becomes too large. The
constraints on the system's size become even more severe if one is
interested in the (nonequilibrium) quantum dynamics of the
system. Feynman proposed the use of a quantum-mechanical simulator to
efficiently solve these types of quantum problems~\cite{feynman}. One
successful platform for simulating lattice spin systems is the trapped
ion quantum simulator, which have already been used to simulate a
variety of scenarios
\cite{two_ion,three_ion,scaling,science,digital,spectroscopy,quench1,quench2}.
In one realization \cite{PorrasCirac}, ions are cooled in a trap to
form a regular array known as a Coulomb crystal and the quantum state
of each simulated electron spin can be encoded in the internal states
of each trapped ion.  Laser illumination of the entire crystal then
can be used to program the simulation (spin-spin interactions,
magnetic fields, etc.) via coupling to phonon modes, and readout of
the internal ion states at the end of the simulation corresponds to a
projective measurement of each simulated spin on the measurement basis.

To date, the largest number of spins simulated in this type of device
is about 300 ions trapped in an a rotating approximately-triangular
lattice in a Penning trap~\cite{britton}. In that experiment, a
spin-dependent optical dipole force was employed to realize an
Ising-type spin-spin coupling with a tunable power law
behavior. However, the Penning trap simulator was not able to perform
certain desirable tasks such as the adiabatic state preparation of the
ground state of a frustrated magnet because it did not include a
time-dependent transverse magnetic field.  The complexity of the
Penning trap apparatus also creates a barrier to adoption and
therefore does not seem to be as widespread as radio-frequency (RF)
Paul traps.

Experiments in linear Paul traps have already performed a wide range
of different quantum simulations within a one-dimensional linear
crystal~\cite{two_ion,three_ion,scaling,science,digital,spectroscopy,quench1,quench2}. The
linear Paul trap is a mature architecture for quantum information
processing, and quantum simulations in linear chains of ions have
benefited from a vast toolbox of techniques that have been developed
over the years.  Initially, the basic protocol \cite{PorrasCirac} was illustrated in a
two-ion trap~\cite{two_ion}, which was followed by a study of the
effects of frustration in a three-ion trap~\cite{three_ion}.  These
experiments were scaled up to larger systems first for the
ferromagnetic case~\cite{scaling} and then for the antiferromagnetic
case~\cite{science}.  Effective spin Hamiltonians were also
investigated using a Trotter-like stroboscopic
approach~\cite{digital}. As it became clear that the adiabatic state
preparation protocol was difficult to achieve in these experiments,
ion trap simulators turned to spectroscopic measurements of excited
states~\cite{spectroscopy} and global quench experiments to examine
Lieb-Robinson bounds and how they change with long-range interactions
(in both Ising and XY models)~\cite{quench1,quench2}.  

It is therefore
desirable to be able to use the demonstrated power of the Paul trap
systems to extend access to the 2D physics that is native to the
Penning trap systems.
However, extension of this technology to higher
dimensions is hampered by the fact that most ions in 2D and 3D Coulomb
crystals no longer sit on the RF null.  This leads to significant
micromotion at the RF frequency and can couple to the control lasers
through Doppler shifts if the micromotion is not perpendicular to the
laser-illumination direction, leading to heating and the congestion of
the spectrum by micromotion sidebands.

In an effort to utilize the desirable features of the Paul trap system
to study the 2D physics,
arrays of tiny Paul traps are being pursued
\cite{Chuang,Sterling,Chiaverini,Kumph,Siverns,Schmied}.  It has also
been shown that effective higher-dimensional models may be programmed
into a simulator with a linear crystal if sufficient control of the
laser fields can be achieved \cite{Korenblit}.
In this article, we study an alternative approach to applying Paul traps
to the simulation of frustrated 2D spin lattices.  We consider a Paul
trap with axial symmetry that forms an oblate potential, squeezing the
ions into a 2D crystal.  The micromotion in this case is purely radial
due to symmetry, and lasers that propagate perpendicular to the
crystal plane will therefore not be sensitive to Doppler shifts from
micromotion.  We study the parameter space of a particular model trap
geometry to find the crystal structures, normal modes, and
programmable spin-spin interactions for 2D triangular crystals in
this trap.  We find a wide parameter space for making such crystals,
and an Ising spin-spin interaction with widely-tunable range, suitable
for studying spin frustration and dynamics on triangular lattices with
tens of ions.

In the future, we expect the simulation of larger systems to be made
possible within either the Penning trap, or in linear Paul traps that
can stably trap large numbers of ions. It is likely that spectroscopy
of energy levels will continue to be examined, including new
theoretical protocols~\cite{spectroscopy_us}. Designing adiabatic
fast-passage experiments along the lines of what needs to be done for
the nearest-neighbor transverse field Ising model~\cite{adiabatic}
might improve the ability to create complex quantum ground states.
Motional effects of the ions are also interesting, such as tunneling
studies for motion in the quantum regime~\cite{aharonov_bohm}. It is
also possible that novel ideas like time crystals~\cite{wilcek,duan}
could be tested (although designing such experiments might be
extremely difficult).

The organization of the remainder of the paper is as follows: In
Sec. 2, we introduce and model the potential energy and effective
trapping pseudopotential for the oblate Paul trap. In Sec. 3, we
determine the equilibrium positions and the normal modes of the
trapped ions, with a focus on small systems and how the geometry of
the system changes as ions are added in. In Sec. 4, we show
representative numerical results for the equilibrium positions and the
normal modes. We then show numerical results for the effective
spin-spin interactions that can be generated by a spin-dependent
optical dipole force. In Sec. 5, we provide our conclusions and
outlook.

\section{Oblate Paul Trap}
The quantum simulator architecture we study here is based on a Paul
trap with an azimuthally symmetric trapping potential that has
significantly stronger axial confinement than radial confinement,
which we call an ``oblate Paul trap'' since the resulting effective
potential resembles an oblate ellipsoid.  As we show below, this can
create a Coulomb crystal of ions that is a (finite) two-dimensional triangular
lattice. 2D Coulomb crystals in oblate Paul traps have been studied
since the 1980's and were used, for instance, by the NIST Ion Storage
Group~\cite{coulomb_cluster} to study
spectroscopy~\cite{doppfree_spec,single_ion_w}, quantum
jumps~\cite{jump}, laser absorption~\cite{absorption1,
  absorption2} and cooling processes~\cite{cooling}. 2D crystals in
oblate Paul traps have also been studied by other groups both
experimentally \cite{Block} and theoretically
\cite{Schiffer,Bedanov,BulutaKitaoka,Buluta,Clossen}.

\begin{figure}
	\includegraphics[scale=0.77]{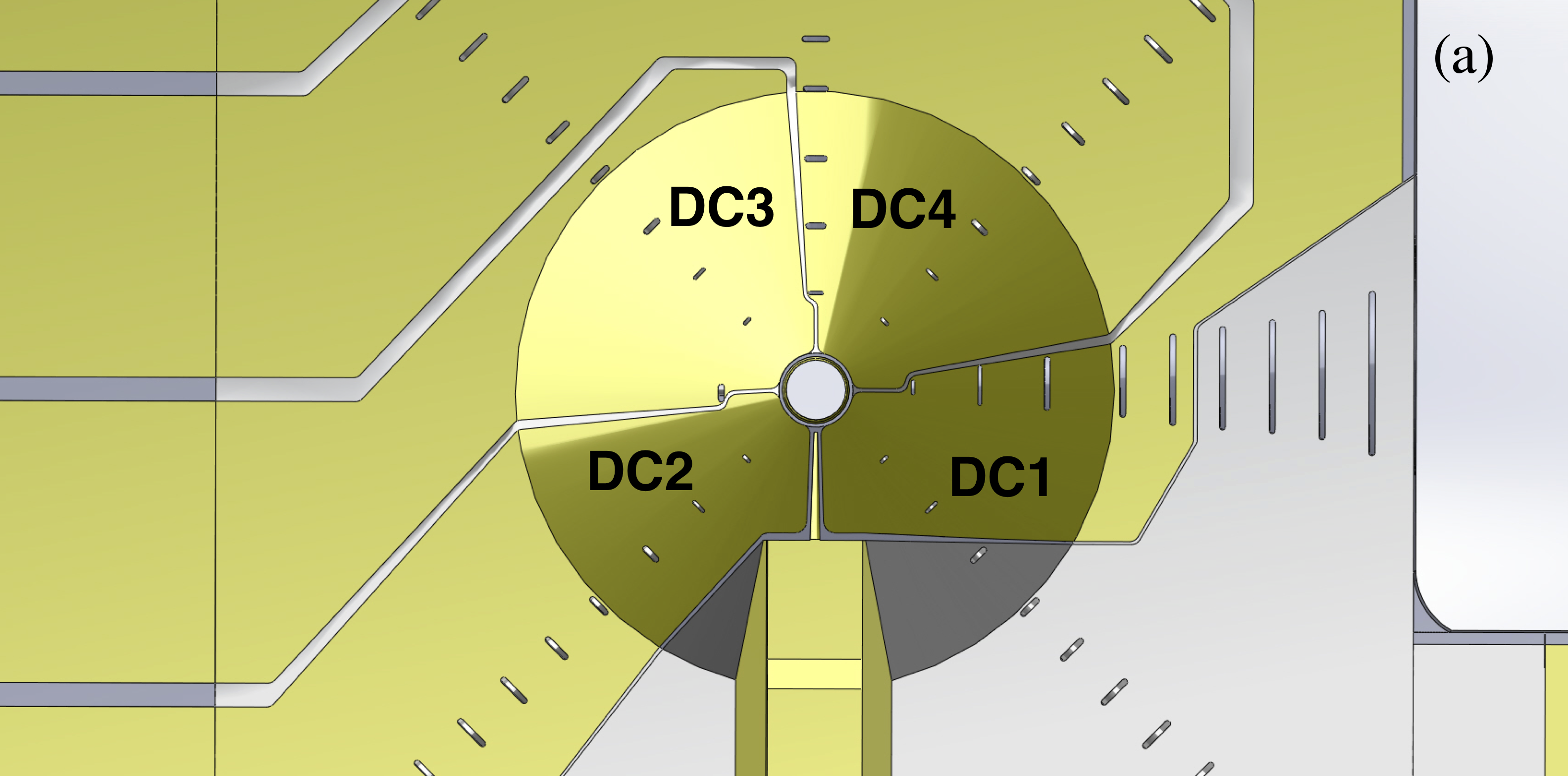} \\
	\includegraphics[scale=0.3]{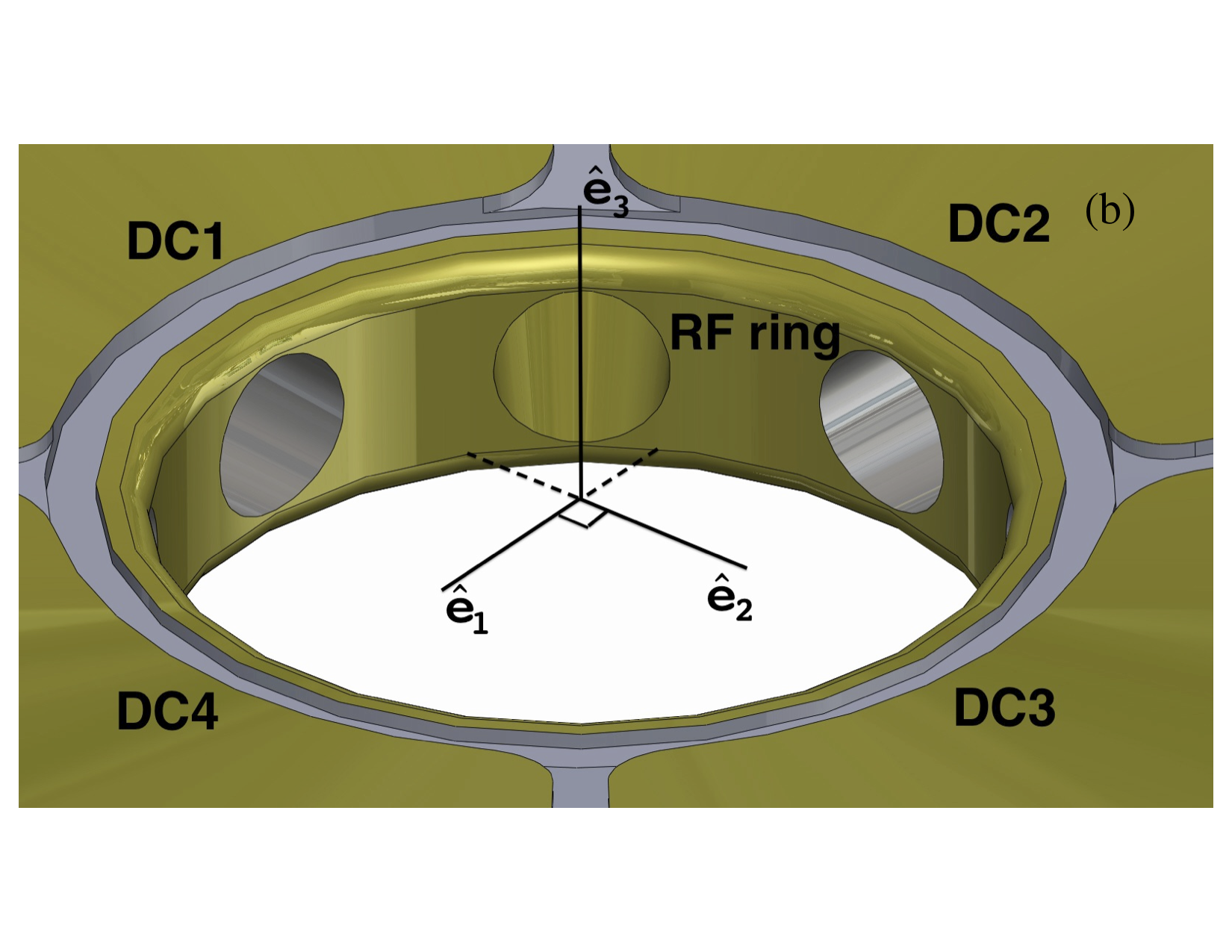}
	\caption{Top view of the proposed oblate Paul trap. The ions
          are trapped in the through hole in the center, which is
          magnified in a three-dimensional image in panel b. Radial
          optical access tunnels are visible in b and will contribute
          to the breaking of rotational symmetry, but play no other
          role in this analysis. Electrodes $1-4$ are labeled in both
          panels and the RF ring is shown in panel b. The origin is
          defined to be at the center of the trap. For our
          calculations, we hold all four electrodes on either the top
          or bottom face at the same potential as the segmenting is
          for compensation of stray fields in the experiment and plays
          no role in the ion crystal structure.}
	\label{fig:draft}
\end{figure}

The particular oblate Paul trap we study has DC ``end cap" electrodes
above and below a central radio-frequency (RF) ring, as depicted in
Fig.~\ref{fig:draft}.  The trap we propose uses modern
microfabrication and lithography technology (manufactured by
Translume, Ann Arbor, MI) to realize the DC end cap electrodes as
surface features on a monolithic fused silica substrate, providing
native mechanical indexing and easier optical access to the ions than
discrete end cap traps. Similar to Penning traps, oblate Paul traps can
be used to study frustration effects when the lattice of ions has
multiple rings. However in a Penning trap, the lattice of ions is
rotating and the ions are in a strong magnetic field, which can add
significant complications. In an oblate Paul trap, the lattice of ions
is stationary except for the micromotion of each ion about its
equilibrium positions (which is confined to the plane of the crystal
by symmetry) and the qubits can be held in nearly zero magnetic field,
permitting the use of the $m=0$ ``clock state" used in linear Paul
trap quantum simulators~\cite{three_ion}. For trapped ions in a
crystal that is a single polygon ($N=3$, $4$, or $5$), we can study
periodic boundary conditions applied to a linear chain of trapped ions
in the linear Paul quantum simulators. Oblate Paul traps can also
potentially be used to perform experiments that are similar to those
recently exploring the Aharonov-Bohm effect~\cite{aharonov_bohm} with
more ions.


It is well known that Maxwell's equations forbid the possibility to
use a static electric field to trap charged particles in free space
through Earnshaw's theorem. However, a static electric field can
create a saddle-point, which confines the charged particles
in some directions and deconfines them in the other directions. A
static electric field that provides a saddle-point is
\begin{equation}
	\mathbf{E}(\tilde{x}_1, \tilde{x}_2, \tilde{x}_3) = A (\tilde{x}_1 \hat{e}_1 + \tilde{x}_2 \hat{e}_2  - 2 \tilde{x}_3 \hat{e}_3),
	\label{eq:Esaddle}
\end{equation}
where $A$ is a constant and $\hat{e}_i$ are the perpendicular unit
vectors with $i = 1,2,3$. Using a static electric field with a
saddle-point, both Penning and  radio-frequency (RF) Paul traps
have successfully trapped charged particles in free space by applying
an additional field. In the Penning trap, one applies a strong uniform
magnetic field, such that the charged particles are confined to a
circular orbit via the Lorentz force, $q\mathbf{v} \times
\mathbf{B}/c$. The RF Paul trap applies a time-varying voltage to its
electrodes, which produce a saddle potential that oscillates
sinusoidally as a function of time. This rapid change of sign allows
for certain ions to be trapped because for particular charge to mass
ratios, the effective focusing force is stronger than the defocussing
force.

\begin{figure}
\centering
	\includegraphics[scale=0.25]{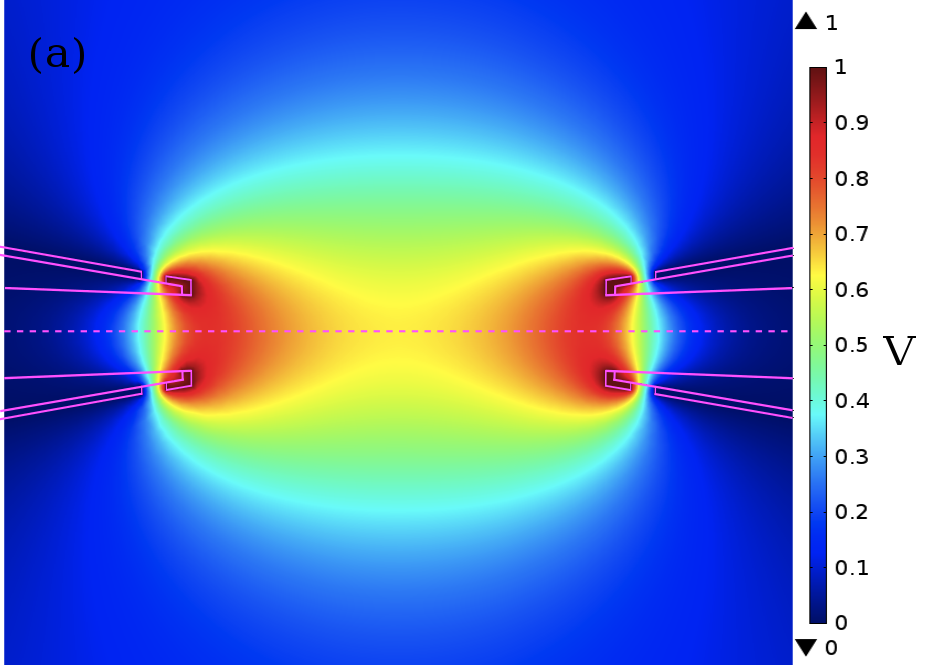} \includegraphics[scale=0.25]{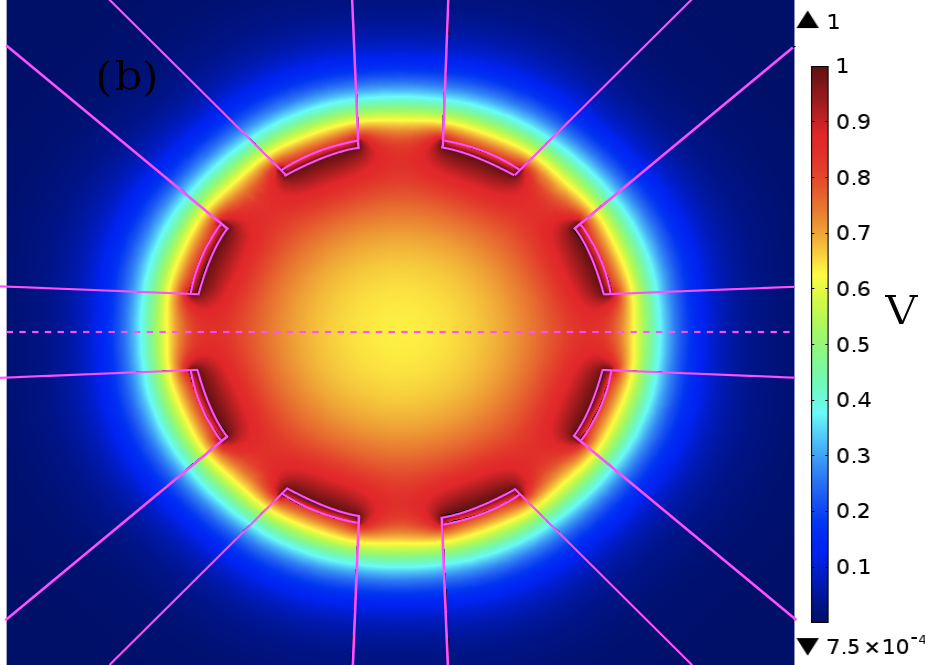}
	\caption{Numerical results for the pseudopotential produced by applying a time-varying voltage with an amplitude of $1$V on the RF ring and grounding the top and bottom electrodes. The oblate Paul trap's edges are shown as solid magenta lines. This numerical result is used to calculate $r_o=512$~$\mu$m. (a) The pseudopotential is shown in the $\hat{e}_1-\hat{e}_3$ plane. The dashed magenta line identifies the plane of panel b. (b) The pseudopotential is shown in the $\hat{e}_1-\hat{e}_2$ plane and the dashed magenta line shows the plane of panel a.}
	\label{fig:RFringSim}
\end{figure}

If the ions remain close to the nulls of the potential, then the
micromotion of the ions is small, and it is a good approximation to
describe the system via a static pseudopotential that approximates the
trapping effect of the time-varying potential.  We use the numerical
modeling software Comsol to simulate this effective pseudopotential
that arises from applying a time-varying voltage to the RF ring and
additional DC voltages on the other electrodes. The effective total
potential energy, $\tilde{V}(\tilde{x}_1, \tilde{x}_2, \tilde{x}_3)$,
of an ion in the oblate Paul trap can be approximated by
\begin{equation}
	\tilde{V}(\tilde{x}_1, \tilde{x}_2, \tilde{x}_3) = \psi(\tilde{x}_1, \tilde{x}_2, \tilde{x}_3) + q\phi(\tilde{x}_1, \tilde{x}_2, \tilde{x}_3),
	\label{eq:SymbTotTrapPot}
\end{equation}
where $\psi(\tilde{x}_1, \tilde{x}_2, \tilde{x}_3)$ is the effective pseudopotential due to the RF fields and $\phi(\tilde{x}_1, \tilde{x}_2, \tilde{x}_3)$ is the additional potential due to the DC voltage applied on the top and bottom electrodes and the RF ring. The resulting pseudopotential at a certain point in space will depend upon the RF frequency, $\Omega_{RF}$, and the RF electric field amplitude, $E_{o,RF}(\tilde{x}_1, \tilde{x}_2, \tilde{x}_3)$, at that point~\cite{dehmelt}and is given by
\begin{equation}
	\psi(\tilde{x}_1, \tilde{x}_2, \tilde{x}_3) = \frac{q^2 }{4m\Omega^2_{RF}}|{\bf E}_{o,RF}(\tilde{x}_1, \tilde{x}_2,\tilde{x}_3)|^2,
	\label{eq:genpseudopot}
\end{equation}
which depends on the charge, $q$, and the mass, $m$, of the particular ion being trapped. After simulating the field using Comsol, we find that the electric field amplitude from the RF field near the trap center can be approximated by
\begin{equation}
{\bf E}_{o,RF} \approx - \frac{2V_{o,RF}}{r^2_o}(\tilde{x}_1\hat{e}_1 +\tilde{x}_2 \hat{e}_2-2\tilde{x}_3\hat{e}_3),
\end{equation}
where $V_{o,RF}$ is the amplitude of the RF voltage.
Plugging this into Eq.~(\ref{eq:genpseudopot}) yields
\begin{equation}
	\psi(\tilde{x}_1, \tilde{x}_2, \tilde{x}_3) = \frac{q^2 V^2_{o, RF}}{m\Omega^2_{RF}r^4_o}(\tilde{x}^2_1 + \tilde{x}^2_2 + 4 \tilde{x}^2_3),
	\label{eq:pseduopot}
\end{equation}
where $r_o=512$~$\mu$m is a fitting parameter, that is determined by grounding the top and bottom electrodes and numerically modeling the square of the RF electric field amplitude, as shown in Fig.~\ref{fig:RFringSim}. We calculate the DC electric field as having $3$ contributions: one from the voltage applied to the RF ring, $\phi_r(\tilde{x}_1, \tilde{x}_2, \tilde{x}_3)$, one from the voltage applied to the top electrodes, $\phi_t(\tilde{x}_1, \tilde{x}_2, \tilde{x}_3)$ and one from the bottom electrodes, $\phi_{b}(\tilde{x}_1, \tilde{x}_2, \tilde{x}_3)$. The DC voltage on the RF ring, $\phi_r(\tilde{x}_1, \tilde{x}_2, \tilde{x}_3)$, is given by
\begin{equation}
	\phi_r(\tilde{x}_1, \tilde{x}_2, \tilde{x}_3) = \frac{V_r}{r^2_o}(\tilde{x}^2_1 +\tilde{x}^2_2 - 2 \tilde{x}^2_3)
	\label{eq:phir}
\end{equation}
where $V_r$ is the DC voltage on the ring.

\begin{figure}[ht!]
\centering
	\includegraphics[scale=0.25]{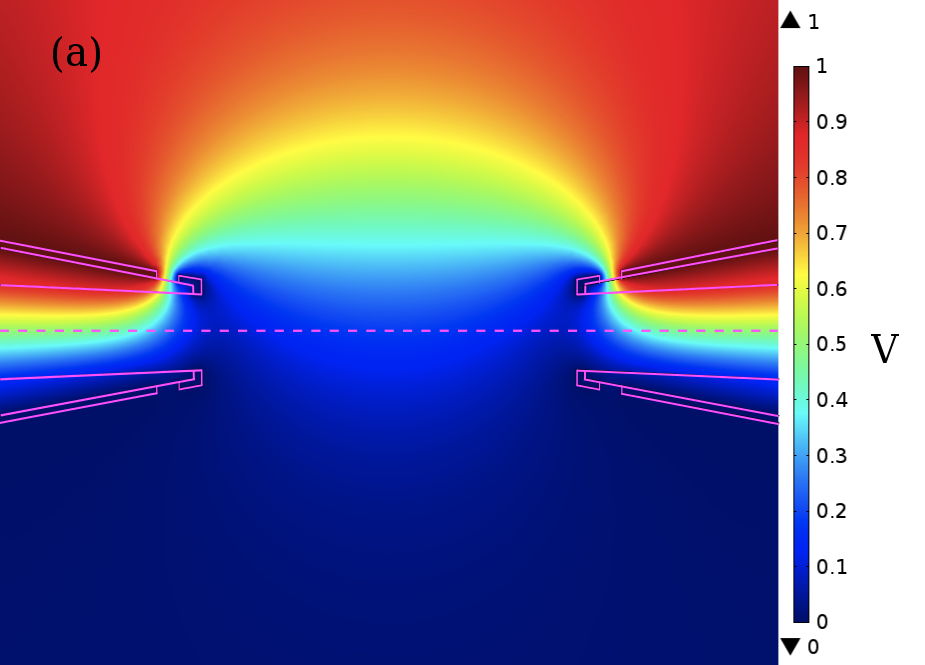} \includegraphics[scale=0.25]{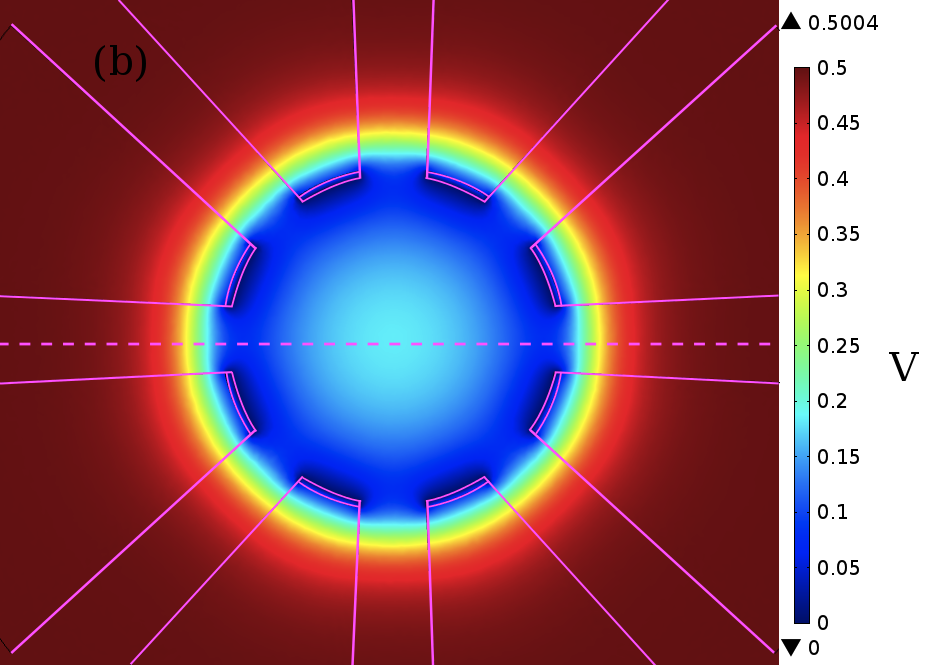}
	\caption{Grounding the RF ring and applying $1$V to the top
          electrodes, we plot the numerical electric potential. The
          oblate Paul trap's electrodes are shown as the solid magenta
          lines. (a) In the $\hat{e}_1-\hat{e}_3$ plane the
          electrostatic potential due to the top electrodes has a
          gradient along the $\hat{e}_3$ direction. This gradient
          produces an electrostatic force that confines the ions in
          the center of the trap. (The dashed magenta line shows the
          plane of panel b.) (b) The electrostatic potential is shown
          in the $\hat{e}_1-\hat{e}_2$ plane, where the the dashed
          magenta line shows the plane of panel a. }
	\label{fig:TBElecSim}  
\end{figure}

We numerically model the electrostatic potential due to the DC voltage applied to the either the top or bottom electrodes, as shown in Fig.~\ref{fig:TBElecSim}. We find that near the trap center, the numerical results for the electrostatic potential produced by a voltage of $V_{t,b}$ the top or bottom electrodes is reasonably modeled by the polynomial
\begin{equation}
\phi_{t,b}(\tilde{x}_1, \tilde{x}_2, \tilde{x}_3) = V_{t, b}\left( \frac{\tilde{x}^2_3}{a^2} + \frac{\tilde{x}_3}{b_{t,b}} - \frac{\tilde{x}^2_1 + \tilde{x}^2_2}{c^2} +d \right),
\label{eq:phitb}
\end{equation}
with fitting parameters $a = 524$~$\mu$m, $b_t = 761$~$\mu$m,
$b_b=-761$~$\mu$m, $c = 704$~$\mu$m, $d=0.812$. Due to the symmetry of
the oblate Paul trap, the parameters satisfy $b_b = -b_t$.


We can use the results from Eqs.~(\ref{eq:pseduopot}-\ref{eq:phitb}) in Eq.~(\ref{eq:SymbTotTrapPot}) to yield the final effective potential energy of an ion in this oblate Paul trap
\begin{eqnarray}
\label{eq:TotTrapPot}
\tilde{V}(\tilde{x}_1, \tilde{x}_2, \tilde{x}_3) &=& \frac{q^2 V^2_{o, RF}}{m\Omega^2_{RF}r^4_o}(\tilde{x}^2_1 + \tilde{x}^2_2 + 4 \tilde{x}^2_3) + q\frac{V_r}{r^2_o}( \tilde{x}^2_1 + \tilde{x}^2_2 - 2 \tilde{x}^2_3)  \\
 &+& qV_t\left( \frac{\tilde{x}^2_3}{a^2} + \frac{\tilde{x}_3}{b_t} - \frac{\tilde{x}^2_1 + \tilde{x}^2_2}{c^2} + d \right) + qV_b\left( \frac{\tilde{x}^2_3}{a^2} + \frac{\tilde{x}_3}{b_b} - \frac{\tilde{x}^2_1 + \tilde{x}^2_2}{c^2} + d\right).\nonumber
\end{eqnarray}
Since the effective potential energy is just a function of $\tilde
x_1^2+\tilde x_2^2$, it is rotationally symmetric around the
$\hat{e}_3$-axis and we would expect there to be a zero frequency
rotational mode in the phonon eigenvectors. That mode can be lifted
from zero by breaking the symmetry, which can occur by adding
additional fields that do not respect the cylindrical symmetry, and
probably occur naturally due to imperfections in the trap, the optical
access ports, stray fields, etc.

\section{Equilibrium structure and normal modes}

Using Eq.~(\ref{eq:TotTrapPot}) (the calculated pseudopotential), we
solve for the equilibrium structure of the crystal in the standard
way. We first construct an initial trial configuration for the ions
and then minimize the total potential energy of the oblate Paul trap
(including the trap potential and the Coulomb repulsion between ions),
as summarized in Eq.~(\ref{eq:V}); MatLab is used for the nonlinear
minimization with a multidimensional Newton's method. We rewrite the
total potential energy of the oblate Paul trap in a conventional form (up
to a constant) via
\begin{eqnarray}
 \label{eq:V}
\tilde{V}(\tilde{x}_1, \tilde{x}_2, \tilde{x}_3)&=&\frac{1}{2}m\left[\sum_{i=1}^{2} (\omega_{\psi, i}^2 + \omega_{r,i}^2 - \omega_{t,i}^2 - \omega_{b,i}^2)\tilde{x}_i^2  +  \omega_{\psi,3}^2 \tilde{x}_3^2 - \omega_{r,3}^2\tilde{x}_3^2\right. \\
 &+& \left. \omega_{t,3}^2 \left( \tilde{x}_3 + \frac{a^2}{2b_t}\right)^2 + \omega_{b,3}^2 \left(\tilde{x}_3 + \frac{a^2}{2b_b}\right)^2\right]
 + \frac{1}{2} \sum_{\substack{m,n=1 \\ m \neq n}}^N 
\frac{k_ee^2}{\tilde{r}_{nm}} ,
\nonumber
\end{eqnarray}
where $k_e= 1/4\pi\epsilon_o$. Here $\tilde{x}_{in}$ is the $i^{th}$
component of the $n^{th}$ ion's location and
$\tilde{r}_{nm}=\sqrt{(\tilde{x}_{1n}-\tilde{x}_{1m})^2 +
  (\tilde{x}_{2n}-\tilde{x}_{2m})^2 +
  (\tilde{x}_{3n}-\tilde{x}_{3m})^2}$. The frequencies in
Eq.~(\ref{eq:V}) are defined via
\begin{equation}
\omega _{\psi, 1} = \frac{\sqrt{2} qV_{0,RF}}{m\Omega _{RF}r_0^2} \text{, }\quad \omega_{\psi, 1} = \omega_{\psi, 2} = \frac{\omega_{\psi, 3}}{2}
\end{equation}
\begin{equation}
\omega_{r, 1} = \sqrt{\frac{2qV_r}{mr_0^2}} \text{, }\quad \omega_{r, 1} = \omega_{r, 2} = \frac{\omega_{r, 3}}{\sqrt{2}}
\end{equation}
\begin{equation}
\omega_{t, 1} = \sqrt{\frac{2qV_{t}}{mc^2}}\text{, }\quad \omega_{et, 1} = \omega_{t, 2} = \frac{a}{c}\omega_{t, 3}.
\label{eq:Omegas}
\end{equation}
We will express all distances in terms of a characteristic length, $l_o$, which satisfies
\begin{equation}
l^3_o = \frac{k_ee^2}{m\omega^2_{\psi,3}}
	\label{eq:natleg}
\end{equation} 
and we will work with dimensionless coordinates $x = \tilde{x}/l_o$
when calculating the equilibrium positions.  Furthermore, we measure
all frequencies relative to $\omega_{\psi,3}$. The normalized
frequencies are
\begin{equation}
 \beta_i = \sqrt{(\omega_{\psi,i}^2 + \omega_{r,i}^2 - \omega_{t,i}^2 - \omega_{b,i}^2)} / \omega_{\psi,3}
\label{eq:beta}
\end{equation}
for $ i = 1, 2$, $\beta_{r,3} = \omega_{r,3}/\omega_{\psi,3}$, $\beta_{t,3} = \omega_{t,3}/\omega_{\psi,3}$, and $\beta_{b,3} = \omega_{b,3}/\omega_{\psi,3}$. The dimensionless total potential energy becomes
\begin{eqnarray}
V = \frac{\tilde{V}}{k_ee^2/l_o} = \frac{1}{2} \sum_{n=1}^N \left[\beta_1^2 x_{1n}^2 + \beta_2^2 x_{2n}^2 + x_{3n}^2 - \beta_{r,3}^2x_{3n}^2 + \right.\\
\left. \beta_{t,3}^2 (x_{3n} + x_{o,t})^2 + \beta_{b,3}^2 ( x_{3n} + x_{o,b})^2\right] + \frac{1}{2} \sum_{\substack{m,n=1 \\ m \neq n}}^N \frac{1}{r_{nm}},
\label{eq:unitlessV}
\end{eqnarray}
where we have defined $x_{o,t} = a^2/(2 l_o b_t)$ and $x_{o,b} = a^2/(2l_ob_r)$. 

To find the equilibrium positions, we use the gradient of the total
potential energy and numerically minimize the total potential energy
using a multidimensional Newton's method. The gradient of
Eq.~(\ref{eq:V}) is
\begin{multline}
\vec{\nabla}V = \sum_{i=1}^3 \sum_{m=1}^N \frac{\partial V}{\partial x_{im}} \hat{e}_{im} = \sum_{m=1}^N \Biggr[\sum_{i=1}^2  \hat{e}_{im} \beta_i^2 x_{im} + \hat{e}_{3m}[x_{3m} - \beta_{r,3}^2 x_{3m}  \\
 + \beta_{t,3}^2 (x_{3m} + x_{o,t}) + \beta_{b,3}^2 (x_{3m} + x_{o,b}) ] + \sum_{\substack{n=1 \\ n \neq m}}^N \sum_{i=1}^3 \hat{e}_{im} \frac{x_{in}-x_{im}}{r_{nm}^3}\Biggr] .
\label{eq:delV}
\end{multline}

The force on ion $m$ in the $\hat{i}$ direction will be
$-\vec{\nabla}V\cdot \hat{e_{im}}$. We seek the solution in which all
ions lie in a plane parallel to the $\hat{e}_1 - \hat{e}_2$ plane,
such that $x_{3m} = \overline{x}_3$ for all $m \in [1,N]$.  As a
result of this condition, $x_{3n} =x_{3m}$, and there is no $x_3$
contribution to the Coulomb potential term. The value of
$\overline{x}_3$ is determined by setting the $\hat{e}_{3m}$ term
equal to zero in Eq.~(\ref{eq:delV}) and is given by the condition
\begin{equation}
\overline{x}_3 = \frac{ - \beta_{t,3}^2 x_{o,t} - \beta_{b,3}^2 x_{o,b}}{1-\beta_{r,3}^2 + \beta_{t,3}^2 + \beta_{b,3}^2} \label{eq:x30},
\end{equation}  
Using $\overline{x}_3$, the ion equilibrium positions are numerically
obtained when all $3N$ components of the force on each ion are zero,
which is given by $\vec{\nabla}V|_{equilib.} = 0$.


After the equilibrium positions $\{\overline{x}_{in}, n = 1, \dots, N,
i = 1,2,3\}$ are found, we expand the total potential about the
equilibrium positions up to quadratic order
\begin{equation}
V=V^{(0)}+ \sum_{i=1}^3 \sum_{m=1}^N q_{im} \frac{\partial}{\partial x_{im}} V|_{eq} + \frac{1}{2} \sum_{i,j=1}^3 \sum_{\substack{n=1 \\ m=1}}^N q_{im}q_{jn} \frac{\partial^2 V}{\partial x_{im} \partial x_{jn}}|_{eq}.
\end{equation}
The nonzero terms of the expansion are the zeroth order and the
quadratic terms; the first order term is zero because the equilibrium
position is defined to be where the gradient of the total potential is
zero, however the zeroth term is also neglected since it is a
constant. We calculate the Lagrangian of the trapped ions using the
quadratic term of the Taylor expanded total potential, with $q_{in}$
being the dimensionless displacement from the equilibrium position for
the $n^{th}$ ion in the $i^{th}$ direction. The expanded Lagrangian
becomes
\begin{equation}
L = \frac{1}{2\omega_{\psi,3}^2} \sum_{i=1}^3 \sum_{m=1}^N \dot{q}_{im}^2 - \frac{1}{2} \sum_{i,j=1}^3 \sum_{m,n=1}^N q_{im} K_{mn}^{ij}q_{jn},
\label{eq:lagrangian}
\end{equation}
where $K_{mn}^{ij}$ represents the elements of the effective spring
constant matrices which are given by
\begin{equation}
(i=1,2) \; K^{ii}_{mn} = 
	\begin{cases}	
		\beta_i^2 - \sum_{\substack{n'=1 \\ n' \neq m}}^N [\frac{1}{\overline{r}_{n'm}^{3}} -3\frac{(\overline{x}_{in'}-\overline{x}_{im})^2]}{\overline{r}_{n'm}^{5}} & \text{if } m=n \\
		\frac{1}{\overline{r}_{mn}^{3}}- 3 \frac{(\overline{x}_{in}-\overline{x}_{im})^2} {\overline{r}_{mn}^{5}} & \text{if } m \neq n
	\end{cases}
\end{equation}
\begin{equation}
K^{12}_{mn} = K^{21}_{mn}=
	\begin{cases}	
		3\sum_{\substack{n'=1 \\ n' \neq m}}^N  \frac{(\overline{x}_{1n'}-\overline{x}_{1m}) (\overline{x}_{2n'}-\overline{x}_{2m})} {\overline{r}_{n'm}^{5}} & \text{if } m = n \\
		-3 \frac{(\overline{x}_{1n}-\overline{x}_{1m}) (\overline{x}_{2n}-\overline{x}_{2m})} {\overline{r}_{mn}^{5}} & \text{if } m \neq n 
	\end{cases}
\end{equation}
\begin{equation}
K^{33}_{mn} =
	\begin{cases}
		\beta_3^2-\sum_{\substack{n'=1 \\ n' \neq m}}^N \frac{1}{\overline{r}_{n'm}^3} & \text{if } m = n \\
		\frac{1}{\overline{r}_{mn}^{3}} & \text{if } m \neq n
	\end{cases}
\label{eq:springMatrix}
\end{equation}
where $\beta_3 = \sqrt{1 - \beta_{p,3}^2 + \beta_{et,3}^2 +
  \beta_{eb,3}^2}$ and $\overline{r}_{mn} = \sqrt{
  (\overline{x}_{1m}-\overline{x}_{1n})^2 +
  (\overline{x}_{2m}-\overline{x}_{2n})^2 }$  is the planar
interparticle distance between ions $n$ and $m$.  Note that motion in
the $3$-direction (axial direction) is decoupled from motion in the
$\hat{e}_1-\hat{e}_2$ plane.

After applying the Euler-Lagrange equation to
Eq.~(\ref{eq:lagrangian}) and substituting the eigenvector solution
$q_{im} = Re(b^\alpha_{im}e^{i\omega_\alpha t})$, we are left to solve
the standard eigenvalue equation
\begin{equation}
-b_{im}^\alpha \left(\frac{\omega_\alpha}{\omega_{\psi, 3}}\right)^2 + \sum_{j=1}^3 \sum_{n=1}^N  K_{mj}^{ij} b_{jn}^\alpha=0.
	\label{eq:eigenproblem}
\end{equation}
There are two sets of normal modes: eigenvectors of the $N \times N$
matrix $K^{33}$ yield the ``axial'' modes (those corresponding to
motion perpendicular to the crystal plane) and eigenvectors of the $2N
\times 2N$ matrix $K^{ij}$, $i,j \in [1,2]$, yield the ``planar''
modes (those corresponding to ion motion in the crystal plane).

\section{Results}

\begin{figure}[!ht]
	\includegraphics[scale=0.1]{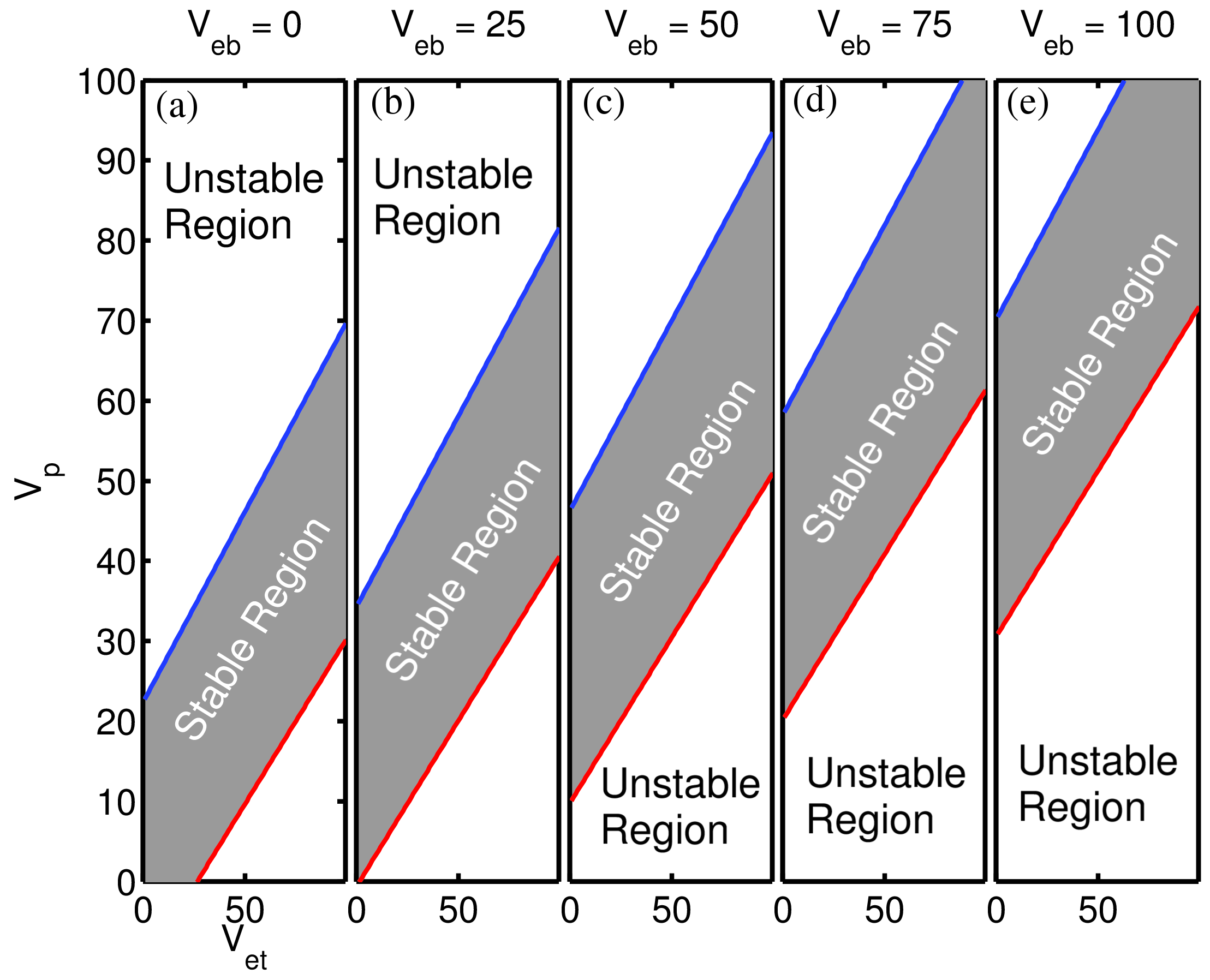}
	\caption{ $\beta_1$ and $\beta_3$ are both real when the ion configuration is stable (shaded grey). We use Eq.~(\ref{eq:beta}) to find a region where the ion configuration is stable. }
	\label{fig:regions}
\end{figure}

Now that we have constructed the formal structure to determine the
equilibrium positions, phonon eigenvectors, and phonon frequencies,
and we have determined the total pseudopotential of the trap, we are
ready to solve these systems of equations to determine the expected
behavior of the trapped ions.  We present several numerical examples
to illustrate the equilibrium structure, eigenvalues of the normal
modes, and the effective spin-spin interaction $J_{mn}$ for the axial
modes with a detuning of the spin-dependent optical dipole force above
the axial center-of-mass phonon frequency, $\omega_{CM}$. We use an
ytterbium ion with mass $m = 171$u, where u is the atomic mass unit,
and a positive charge $q=e$. For the frequency of the RF voltage, we
use $\Omega_{RF} = 2\pi \times 35$MHz and the amplitude of the
potential applied to the RF ring is $V_{o, RF} \approx 500$V. The DC
voltage applied to the RF ring and to the top and bottom electrodes
will be $|V_{DC}| \leq 100$V. We work in a region where the
trapped ion configurations are stable. The ion crystal is stable only when both
$\beta_1$ and $\beta_2$ are real and nonzero, as defined in
Eq.~(\ref{eq:beta}) and this region is shown in
Fig.~\ref{fig:regions}, which depends on the voltages applied to the
RF ring and to the top and bottom electrodes. We work with ion
crystals that contain up to $20$ ions.

\subsection{Equilibrium configurations}
\begin{figure}[!ht]
	\centering
	\begin{tabular}{c c}
		\includegraphics[scale=0.3]{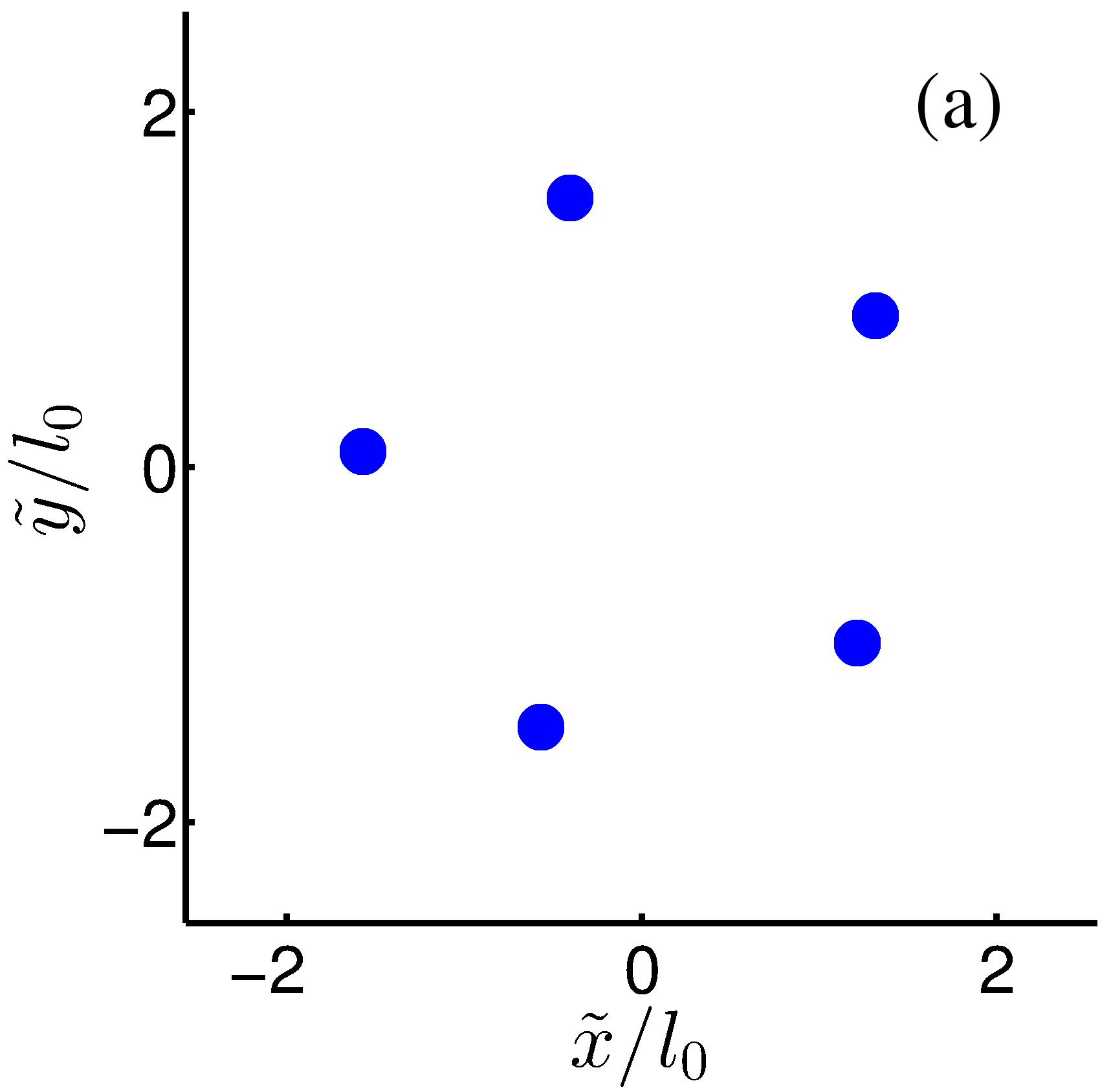} & \includegraphics[scale=0.3]{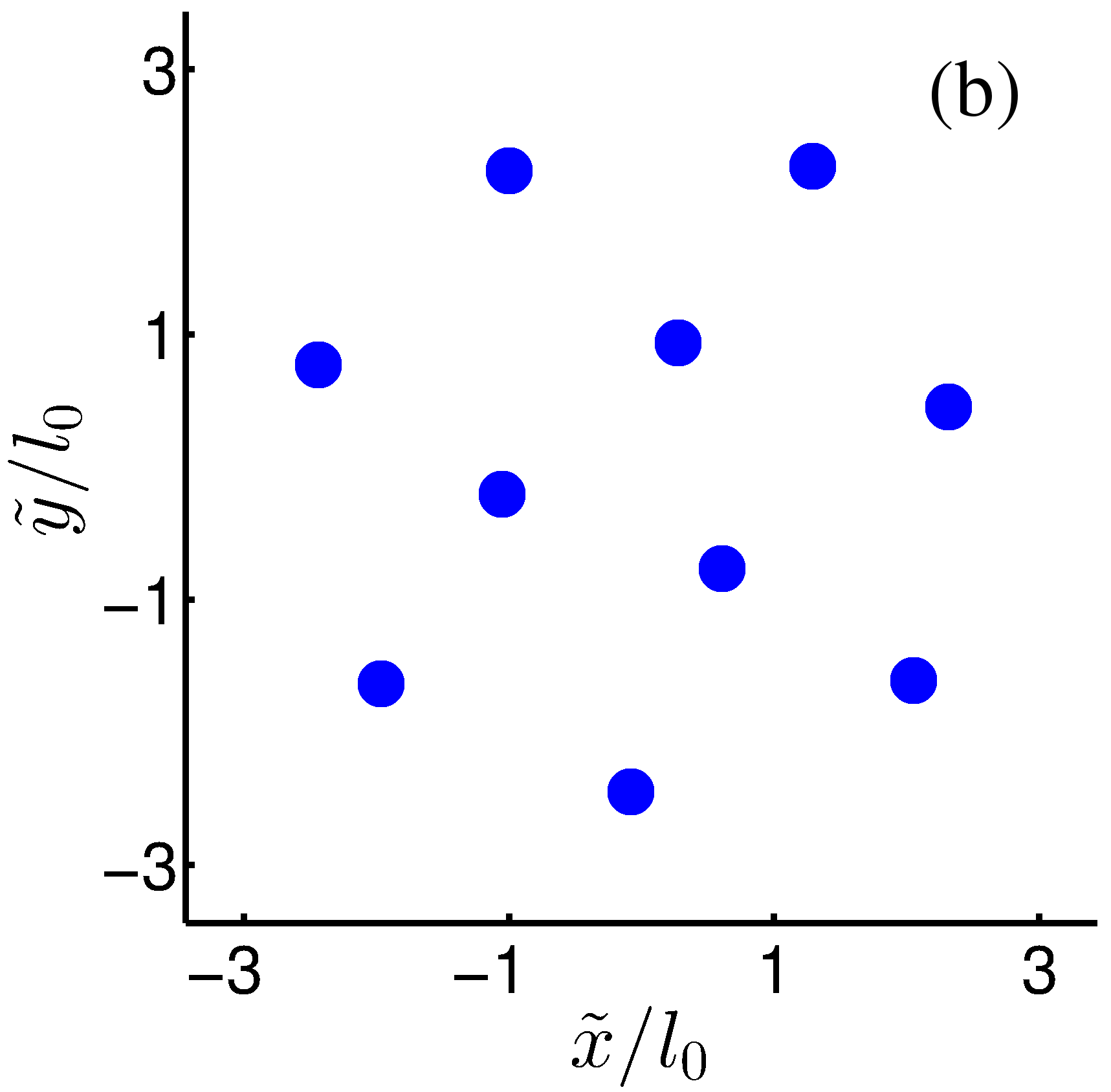} \\
		\includegraphics[scale=0.3]{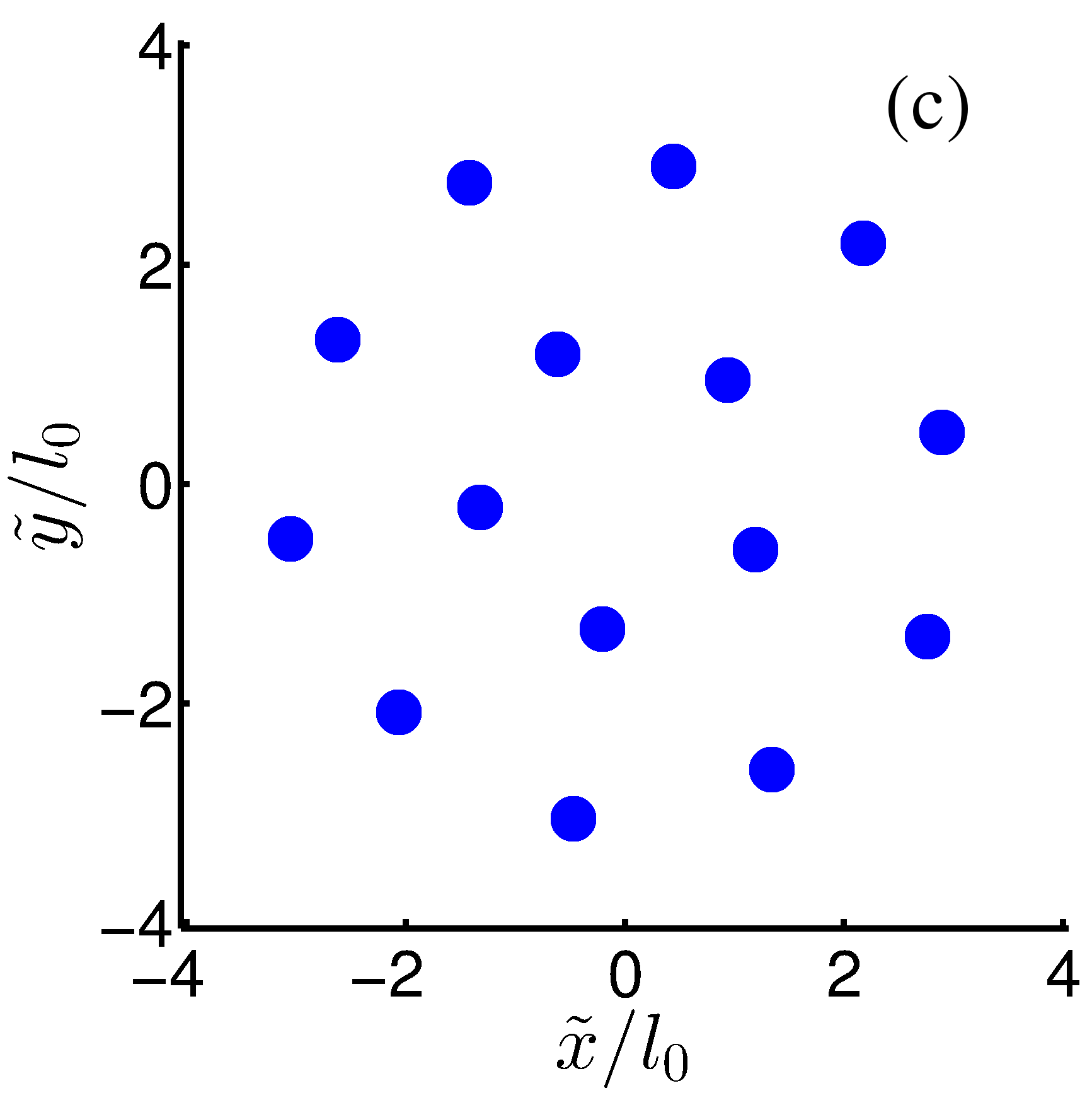} & \includegraphics[scale=0.3]{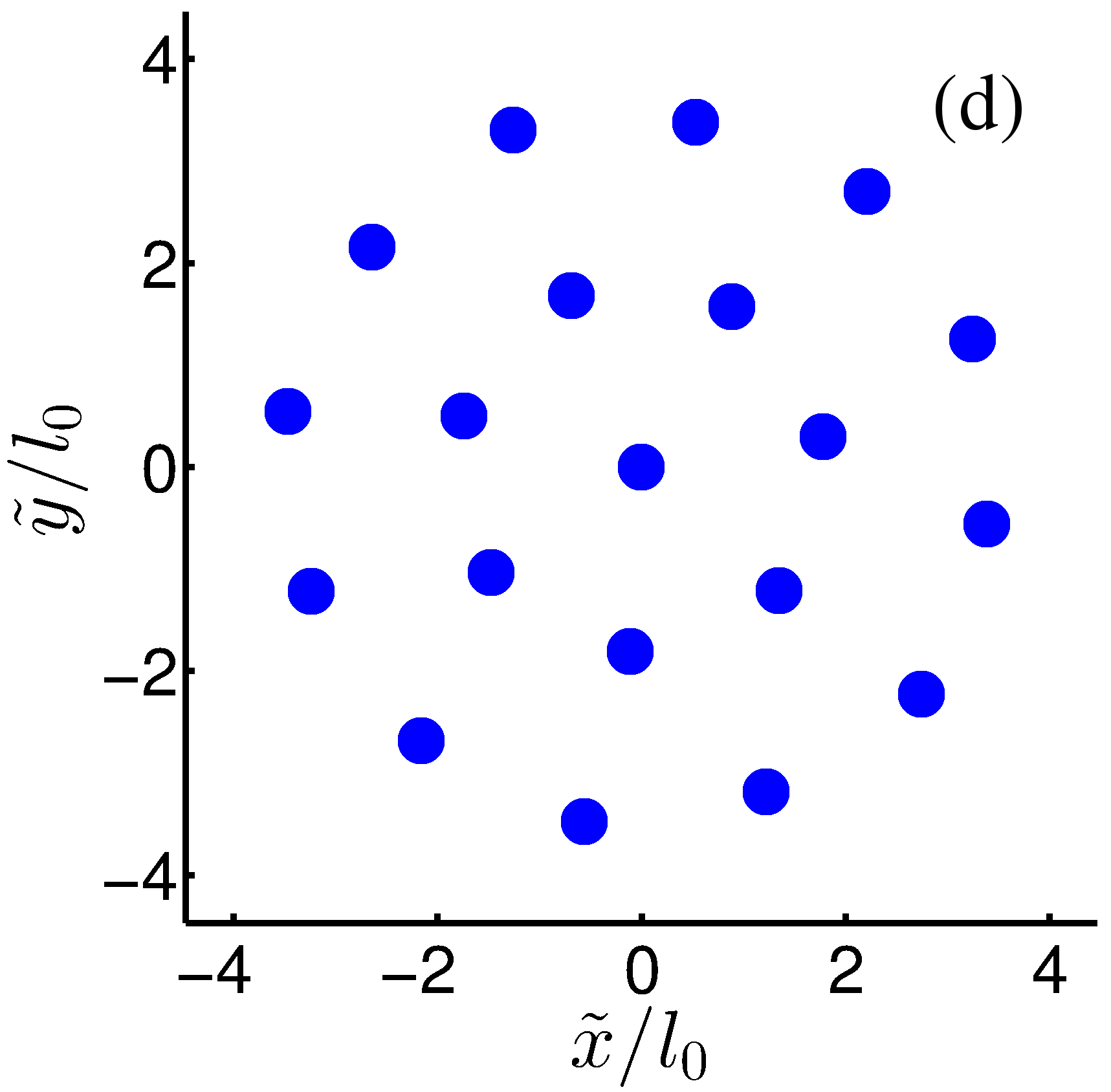}
	\end{tabular}
	\caption{Equilbrium positions (blue dots) calculated for $N =5$ (a), $10$(b), $15$(c), $20$(d) with $V_r=46.3$V and $v_t=V_b=50$V. (a) $N=5$ is the maximum number of ions for a single ring. (b) $N=10$ is the first instance where there is a second ring in the center of the configuration. (c) Similar to panel (a), $N=15$ is the maximum number of ions to have two rings. (d) $N=20$ is the maximum number ions expected to operate within our trap, As previously noted, the additional ions when added to the $N=15$ equilibrium configuration occupy the outer rings instead of the center.}
	\label{fig:EquilibPos}
\end{figure}

A single ion will sit in the center of the trap.  As more ions are added, because the ions repel each other, a ``hard core'' like structure will form, starting with rings of ions until it becomes energetically more stable to have one
ion in the center of the ring, and then additional shells surrounding it, and so on. We find that as we increase the number of ions, the single ring is stable for $N=3$, 4, and 5. Increasing $N$ further creates more complex structures.
We show the common equilibrium configurations for $N=5$, $10$, $15$, $20$ with DC voltages of $V_r = 46.3$V and $V_t=V_b=50$V, in Fig.~\ref{fig:EquilibPos}. As mentioned above, $N=5$ is the last configuration that is comprised of a single ring of ions, as depicted in Fig.~\ref{fig:EquilibPos}(a). The $N=5$ configuration is ideal to use in order to study when periodic boundary conditions are applied to the linear chain, this is due to the configuration being in a single ring. For configurations with $N = 6$ through $8$, the additional ions are added to an outer ring. When $N=9$ the additional ions are added to the center. In Fig.~\ref{fig:EquilibPos}(b), $N=10$ is the first configuration the forms a ring in the center, with three ions. The equilibrium configuration of $N=15$ is the maximum number to have two rings, as shown in Fig.~\ref{fig:EquilibPos}(c). Ion configurations with $N > 15$ have a single ion at the center, as an example of this, we show $N=20$ in Fig.~\ref{fig:EquilibPos}(d). The common configurations for $N > 5 $ are nearly formed from triangular lattices (up to nearest neighbor) and this could be used to study frustration in the effective spin models (except, of course, that due to the finite number of ions there are many cases where the coordination number of an interior ion is not equal to 6, as seen in Fig.~\ref{fig:EquilibPos}). The shape of all of these clusters for small $N$ agree with those found in Ref.~\cite{Bedanov}, except for $N=10$, 12, and 14, which have small differences due to the different
potential that describes the oblate Paul trap from the potential used in~\cite{Bedanov}.

\begin{figure}[!ht]
	\centering
	\begin{tabular}{c c}
		\includegraphics[scale=0.3]{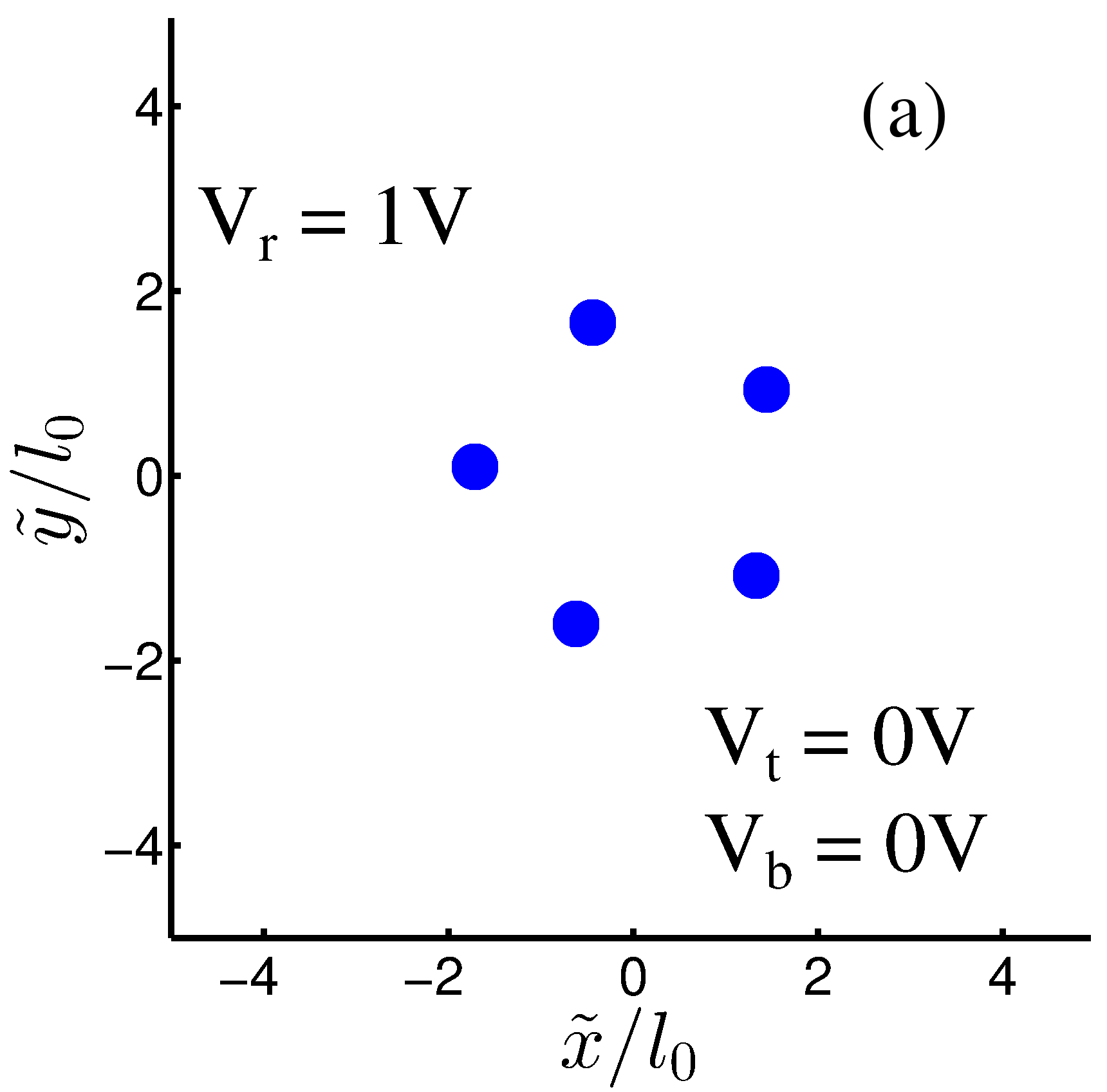} & \includegraphics[scale=0.3]{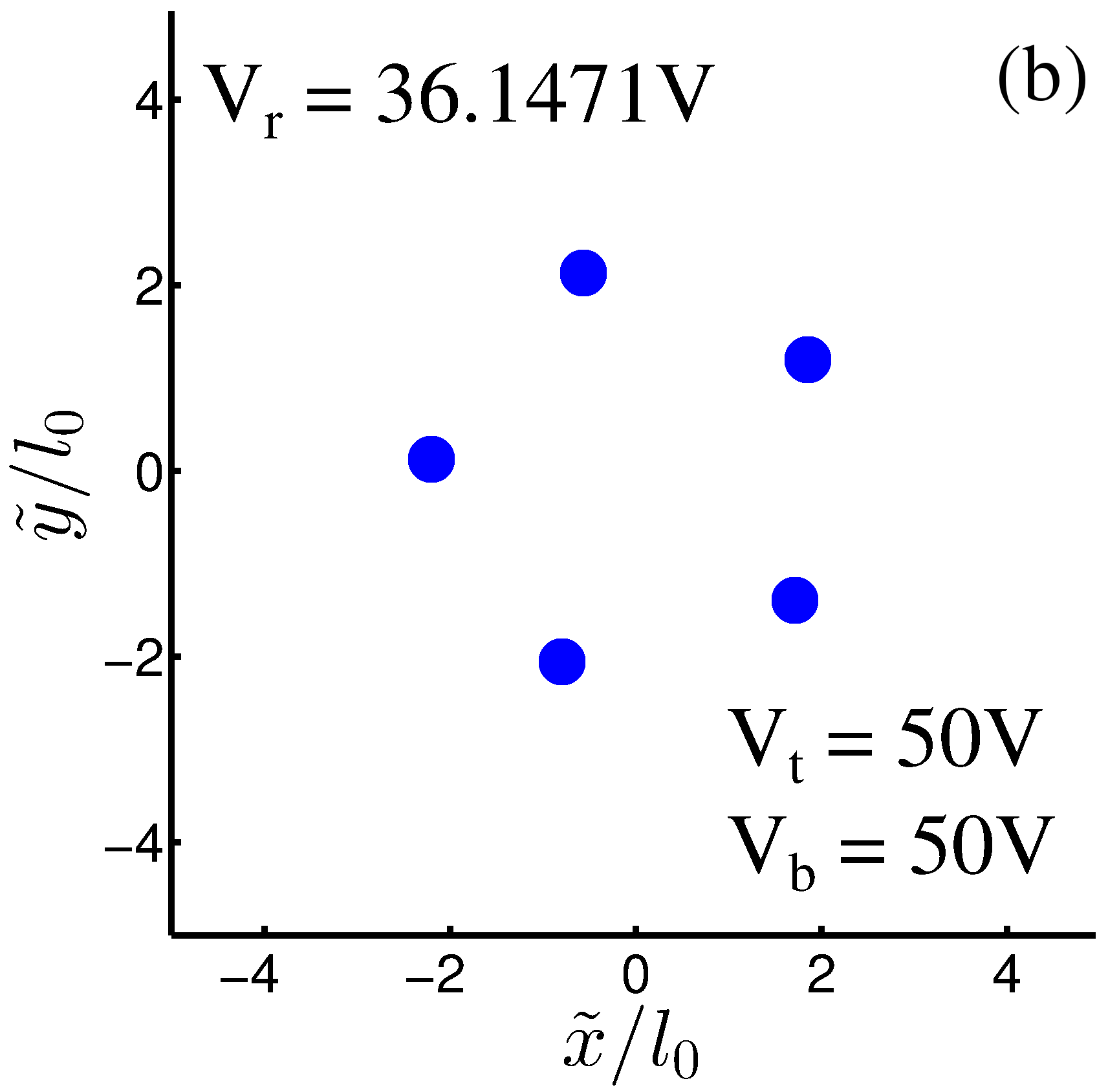} \\
		\includegraphics[scale=0.3]{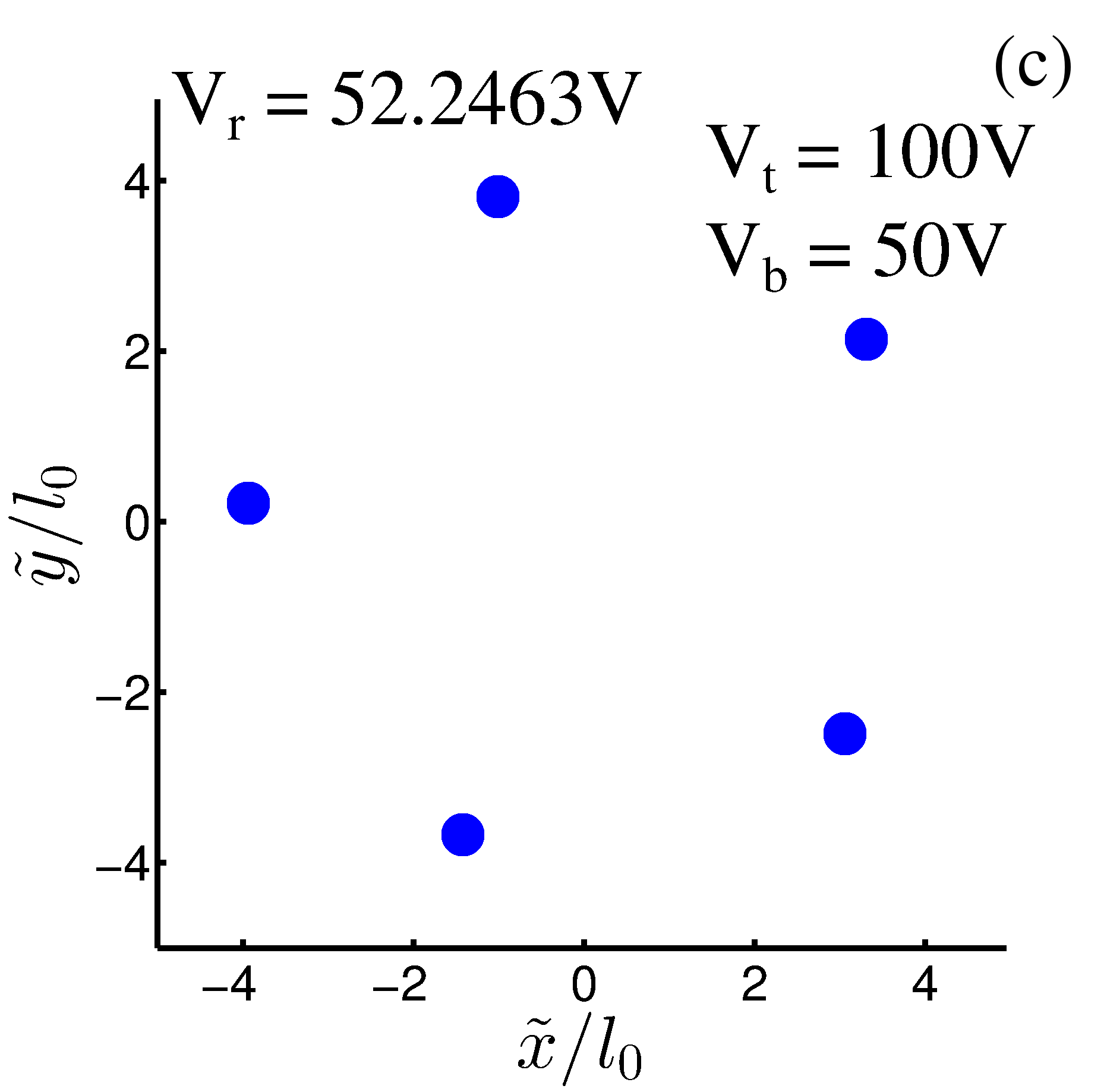} & \includegraphics[scale=0.3]{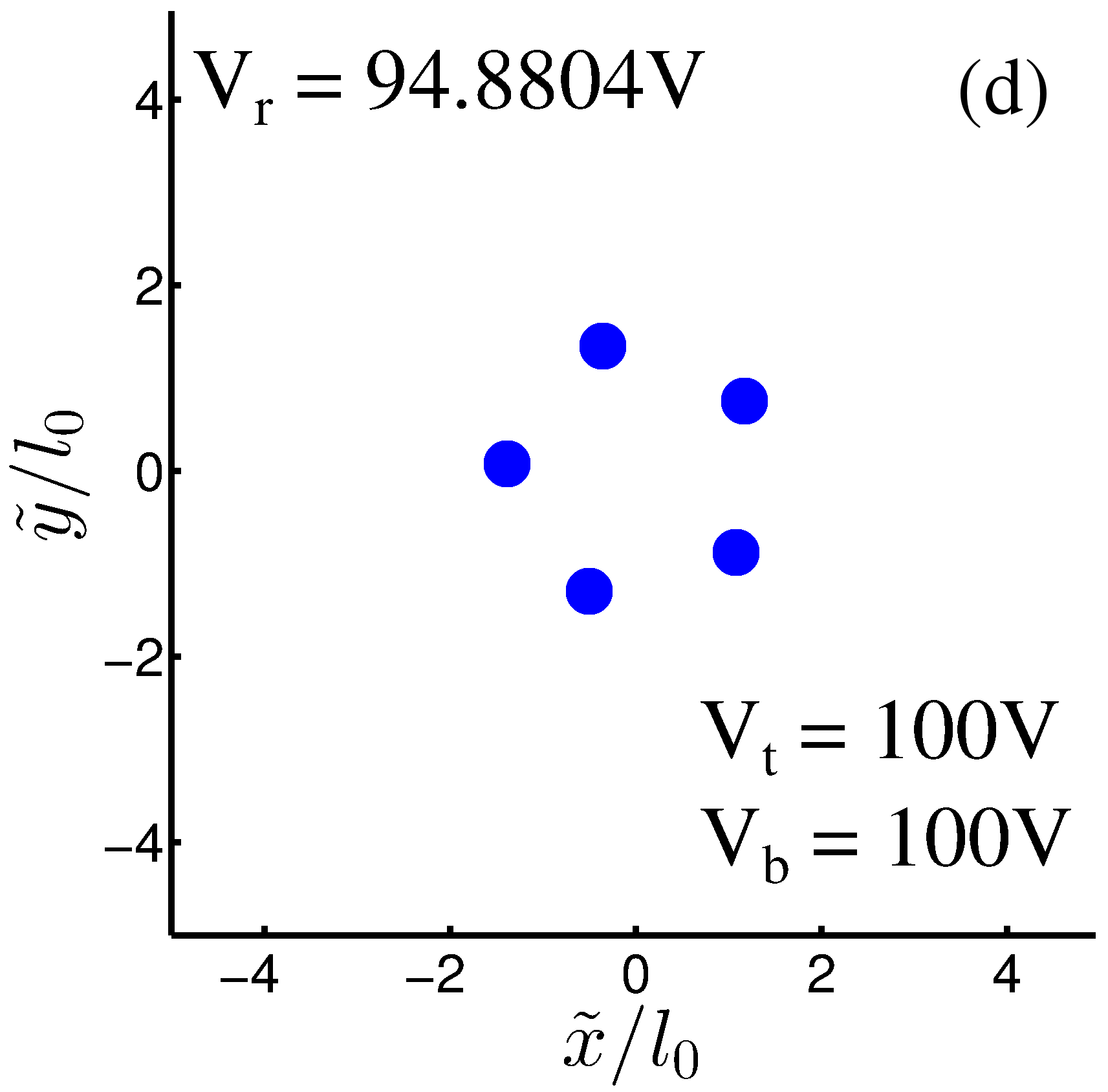}
	\end{tabular}
	\caption{ Equilibrium positions (blue dots) for $N=5$ with various DC voltages on the RF ring and independently applied to the top and bottom electrodes. The equilibrium shape remains the same for the four cases shown: (a) $V_r=1$V and $V_t=V_b= 0$V, (b) $V_r=36.147$V and $V_t=V_b= 50$V, (a) $V_r=52.225$V, $V_t= 100$V, and $V_b= 50$V and (d) $V_r=94.880$V and $V_t=V_b= 100$V.}
	\label{fig:5varyingV}
\end{figure}

We next show the dependence of the equilibrium positions on the DC voltages applied to the RF ring and independently applied to the top and bottom electrodes. We fix $N=5$. As each DC voltage is independently varied, the shape of the equilibrium configuration for $N=5$ remains the same and only the distances between ions change, as shown in the four cases in Fig.~\ref{fig:5varyingV}.  

\subsection{Normal modes}
After determining the equilibrium positions, we can find the spring
constants and then solve the eigenvalue problem to find the normal
modes. Note that due to rotational symmetry, there always is a zero
frequency planar mode corresponding to the free rotation of the
crystal. In an actual experiment, however, we expect that the
rotational symmetry of the trap will be broken by stray fields, the
radial optical access tunnels, imperfections in the electrodes, etc.,
so that mode will be lifted from zero.

\begin{figure}[!ht]
	\centering
	\begin{tabular}{c c}
		\includegraphics[scale=0.3]{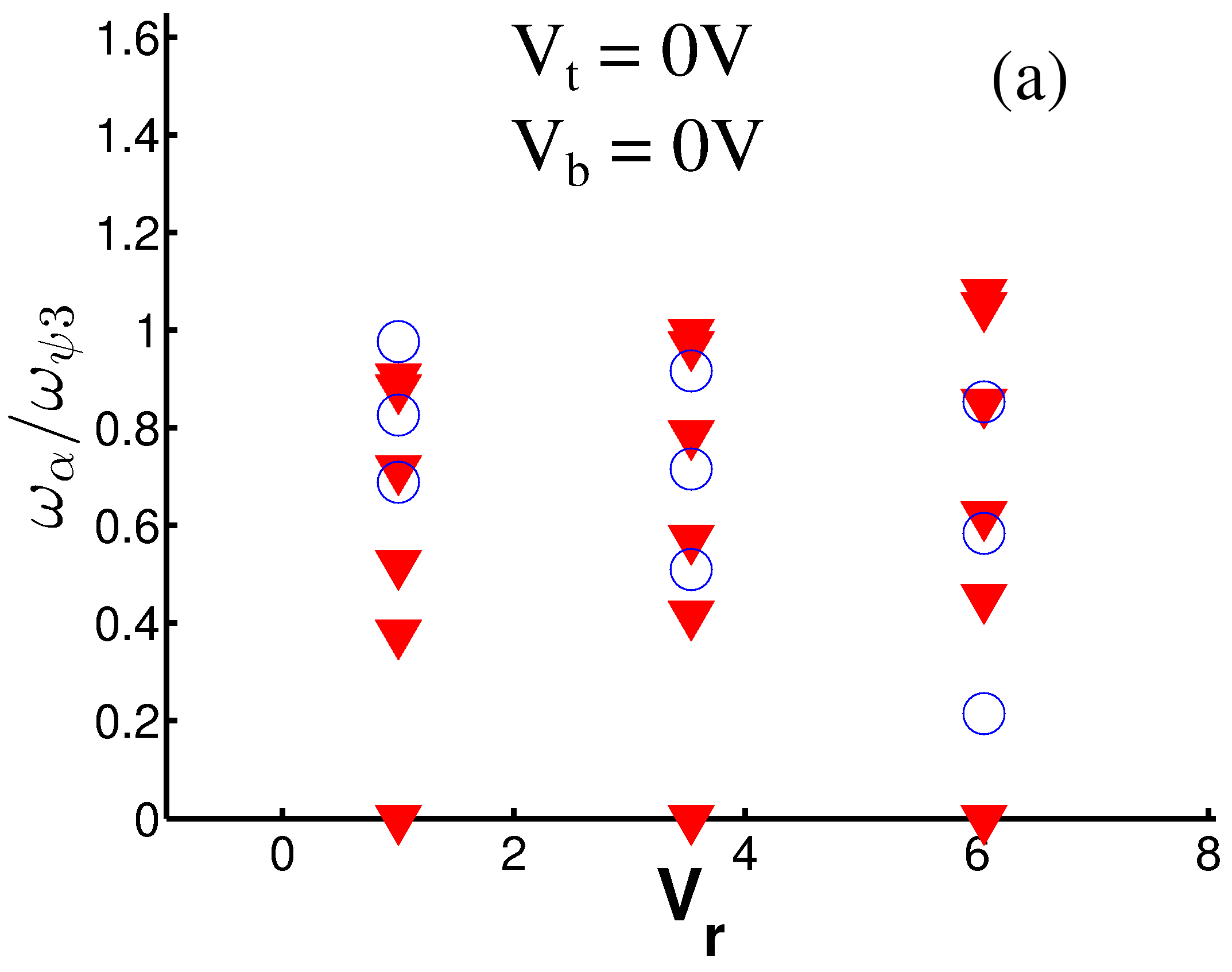} & \includegraphics[scale=0.3]{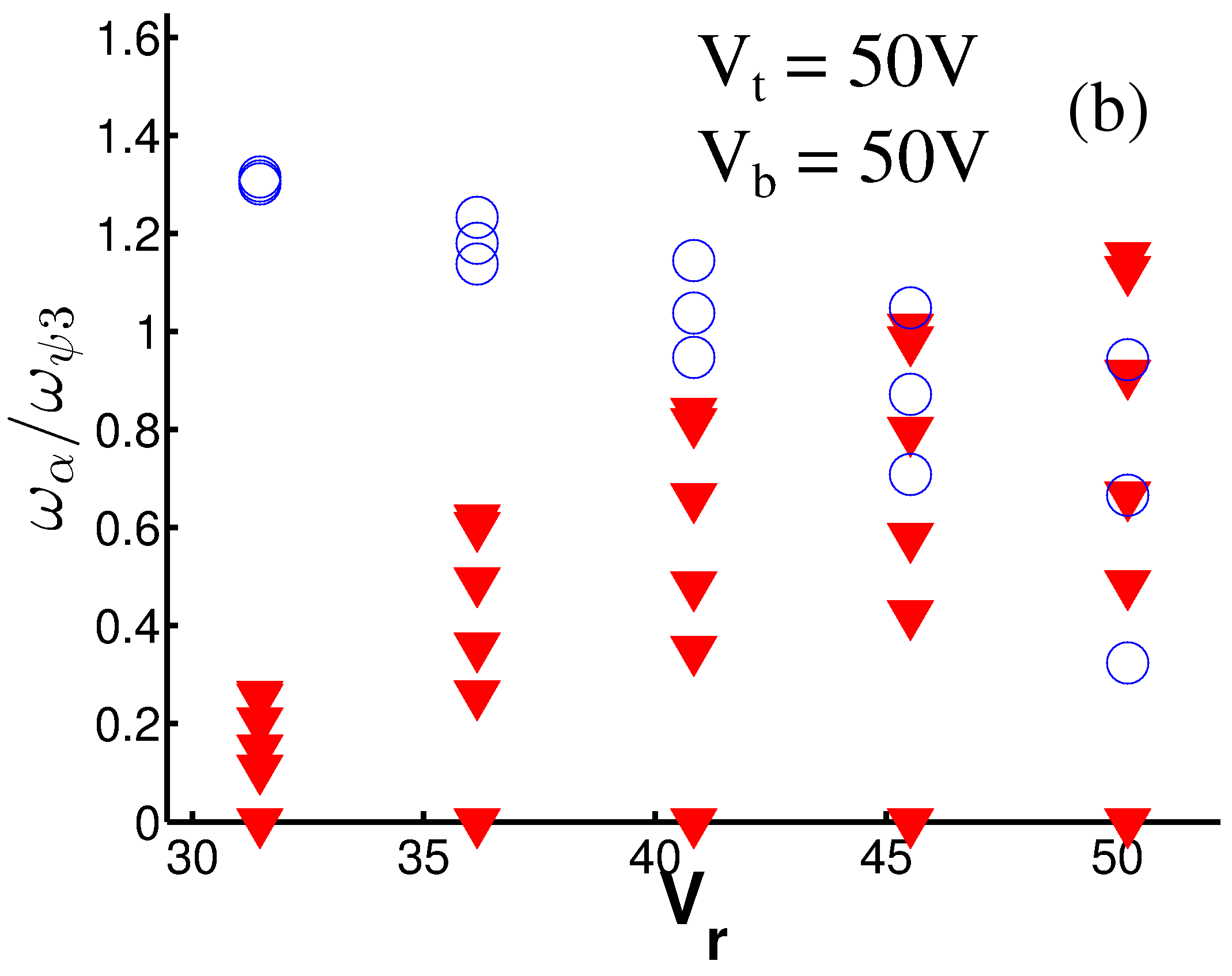} \\
		\includegraphics[scale=0.3]{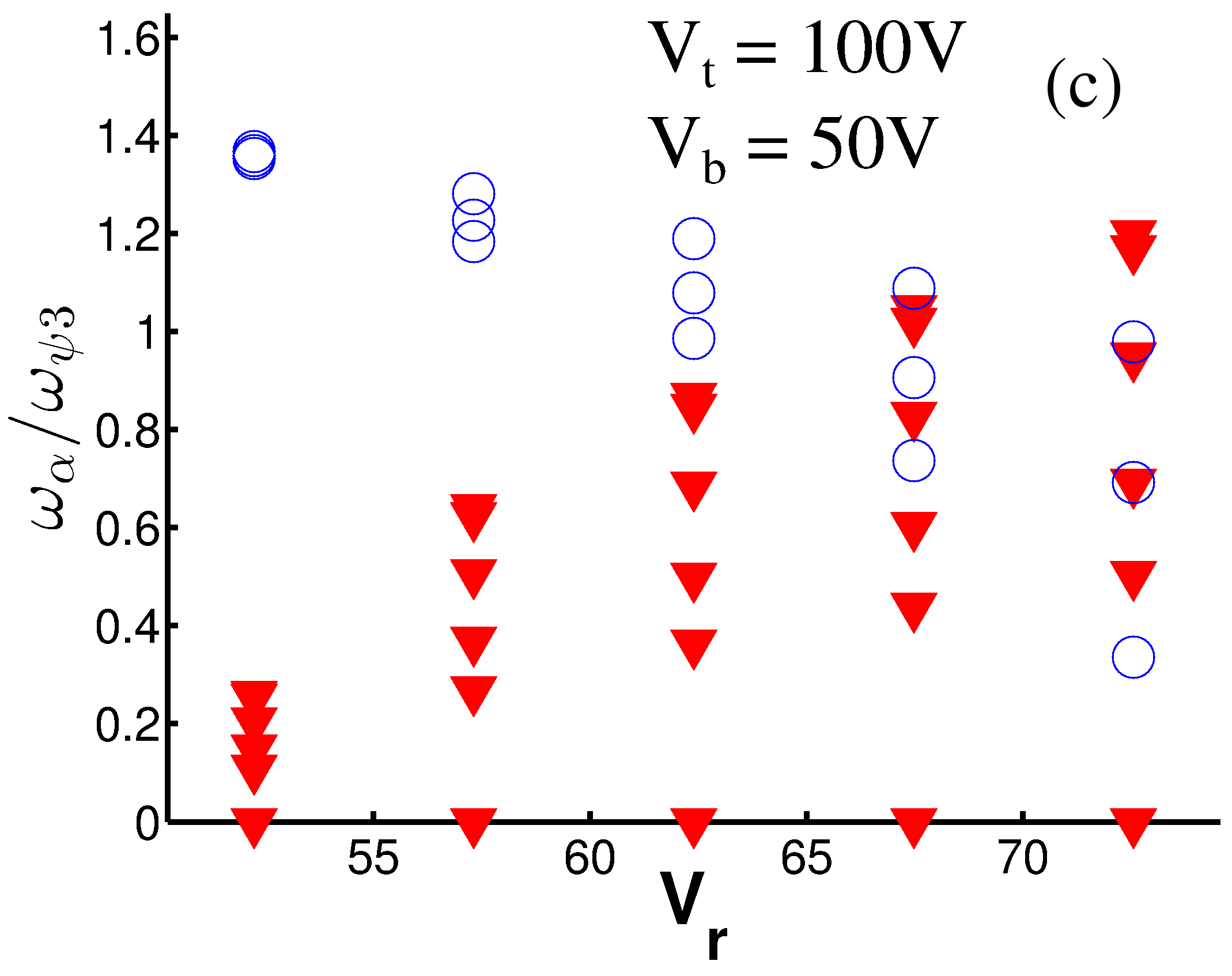} & \includegraphics[scale=0.3]{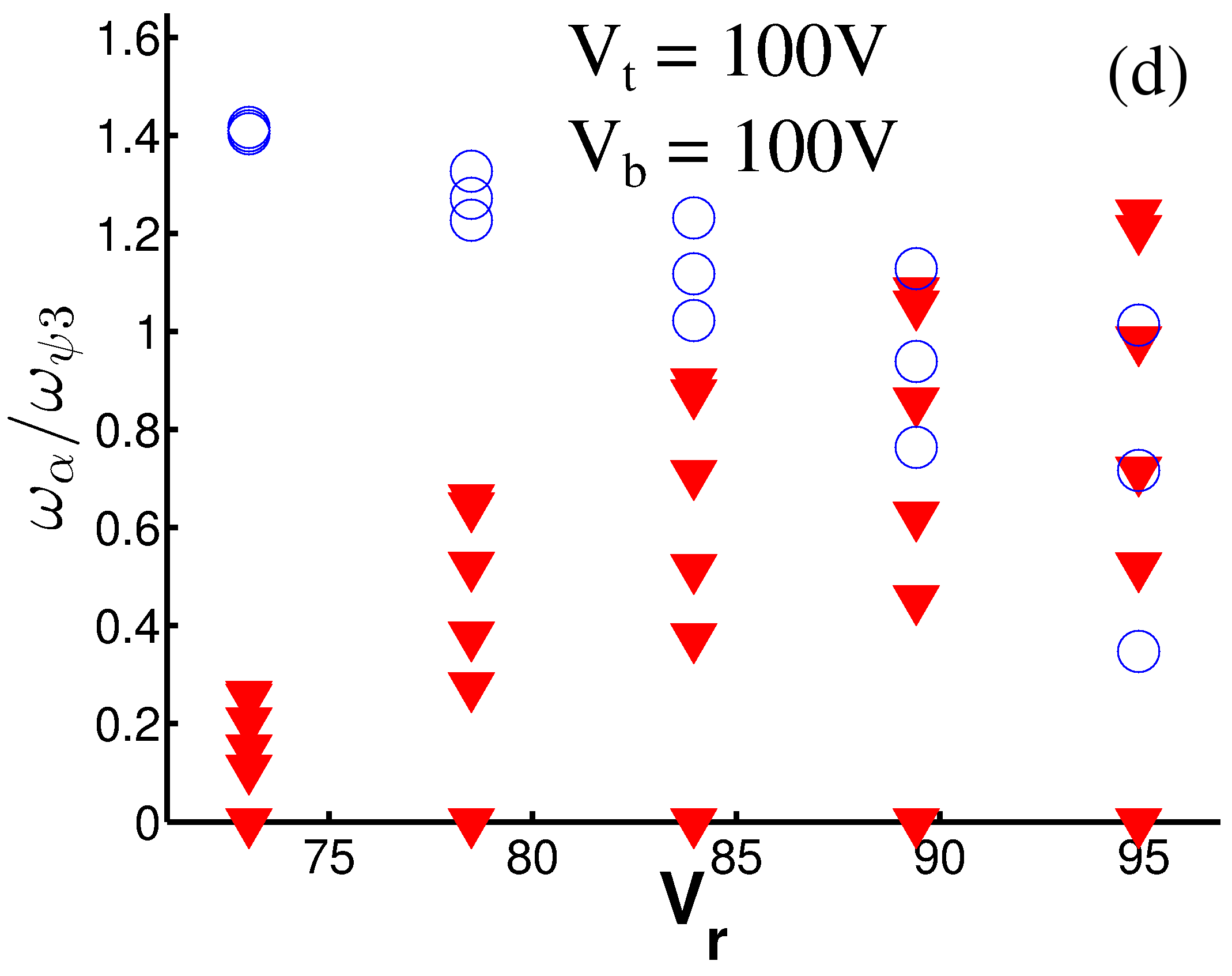}
	\end{tabular}
	\caption{ Eigenvalues of the normal modes as a function of
          $V_r$. The axial mode frequencies (blue circles) decrease
          and planar mode frequencies (red triangles) increase as
          $V_r$ increases, as shown in the the four cases illustrated:
          (a) $V_t=V_b= 0$V, (b) $V_t=V_b= 50$V, (a)$V_t= 100$V and
          $V_b= 50$V and (d) $V_t=V_b= 100$V. For (b-d) the
          eigenvalues of the normal modes also are closer together at
          low $V_r$ and separate as $V_r$ increases. }
	\label{fig:Eigen5}
\end{figure}

We show the eigenvalues of the normal modes for $N=5$ in
Fig.~\ref{fig:Eigen5}. The axial phonon frequencies decrease as the DC
voltage on the RF ring increases and the planar phonon frequencies
decrease as the DC voltage on the RF ring decreases. For the majority
of the combinations of $V_t$ and $V_b$ the axial phonon frequencies
lie in a narrow band that is separated from the planar mode
frequencies, which also lie in a narrow band. As $V_r$ increases the
axial band broadens and eventually overlaps the planar band, which is
also broadening.  When the axial band has an eigenvalue that goes
soft, the system is no longer stable within one plane (which is the
equivalent of the zig-zag transition in the linear Paul trap). When
$V_t=V_b=0$ the initial clustering of the axial modes and planar modes
is not present, as shown in Fig.~\ref{fig:Eigen5}(a).

\subsection{Ising spin-spin interaction}

The ions in our trap have two hyperfine states that are separated by a frequency $\omega_0$.  Three laser beams with two beatnotes at frequencies $\omega_0 \pm \mu$ will illuminate the ions, selectively exciting phonon modes as described in~\cite{three_ion}. In this case, we choose the laser beams to propagate along the $\pm \hat{e}_3$ direction, as defined in Fig.~\ref{fig:draft}b, such that the laser beams are insensitive to the micromotion which is entirely radial. The phonon modes are excited in a spin-dependent way to generate effective spin-spin interactions which can be described by the Ising spin coupling matrix, $J_{mn}$
\begin{equation}
\mathcal{H} = \sum^N_{mn} J_{mn}\sigma^{z}_m\sigma^{z}_n,
\end{equation}
where $\sigma^{z}_i$ is the Pauli spin matrix of ion $i$ in the $\hat{e}_3$-direction and we have neglected the time-dependent terms of the spin couplings $J_{mn}$. The explicit formula for $J_{mn}$ is~\cite{monroe_duan}
\begin{equation}
J_{m,n} = J_0 \sum_\alpha \frac{b^*_{m,\alpha}b_{n,\alpha}}{ (\frac{\mu}{\omega_{CM}})^2 - (\frac{\omega_m}{\omega_{CM}})^2},
\end{equation}
where the coefficient $J_0$, depends on the carrier transition Rabi frequency ($\Omega$), the difference in wavevector between the laser beams ($\delta k$), the ion mass ($m$), and the frequency of the center-of-mass mode, ($\omega_{CM}$), and is given by
\begin{equation*}
J_0 = \frac{\Omega^2 \hbar (\delta k)^2}{2m\omega_{CM}^2}.
\end{equation*} 
 
\begin{figure}[!ht]
	\centering
	\includegraphics[scale=0.325]{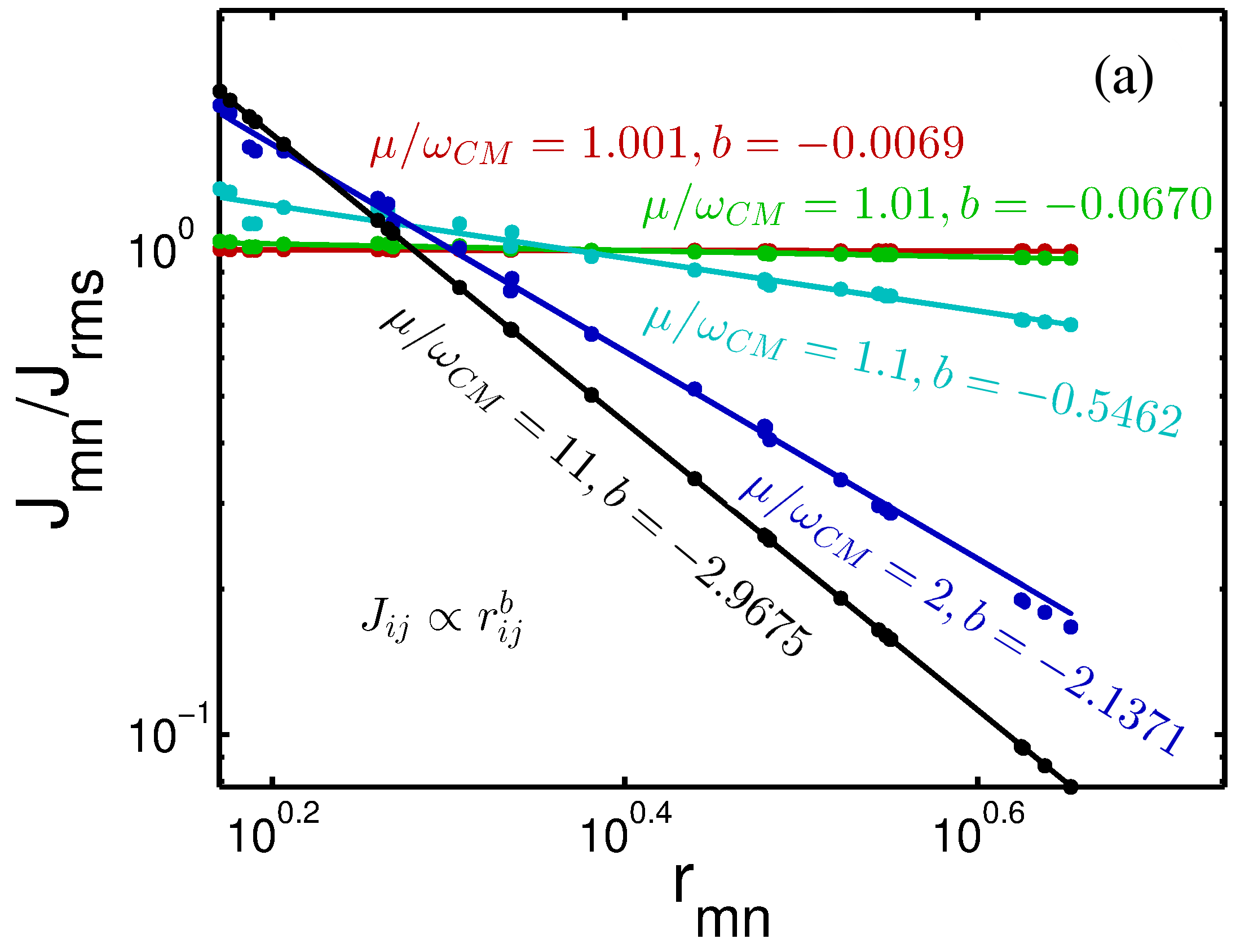}
	\includegraphics[scale=0.325]{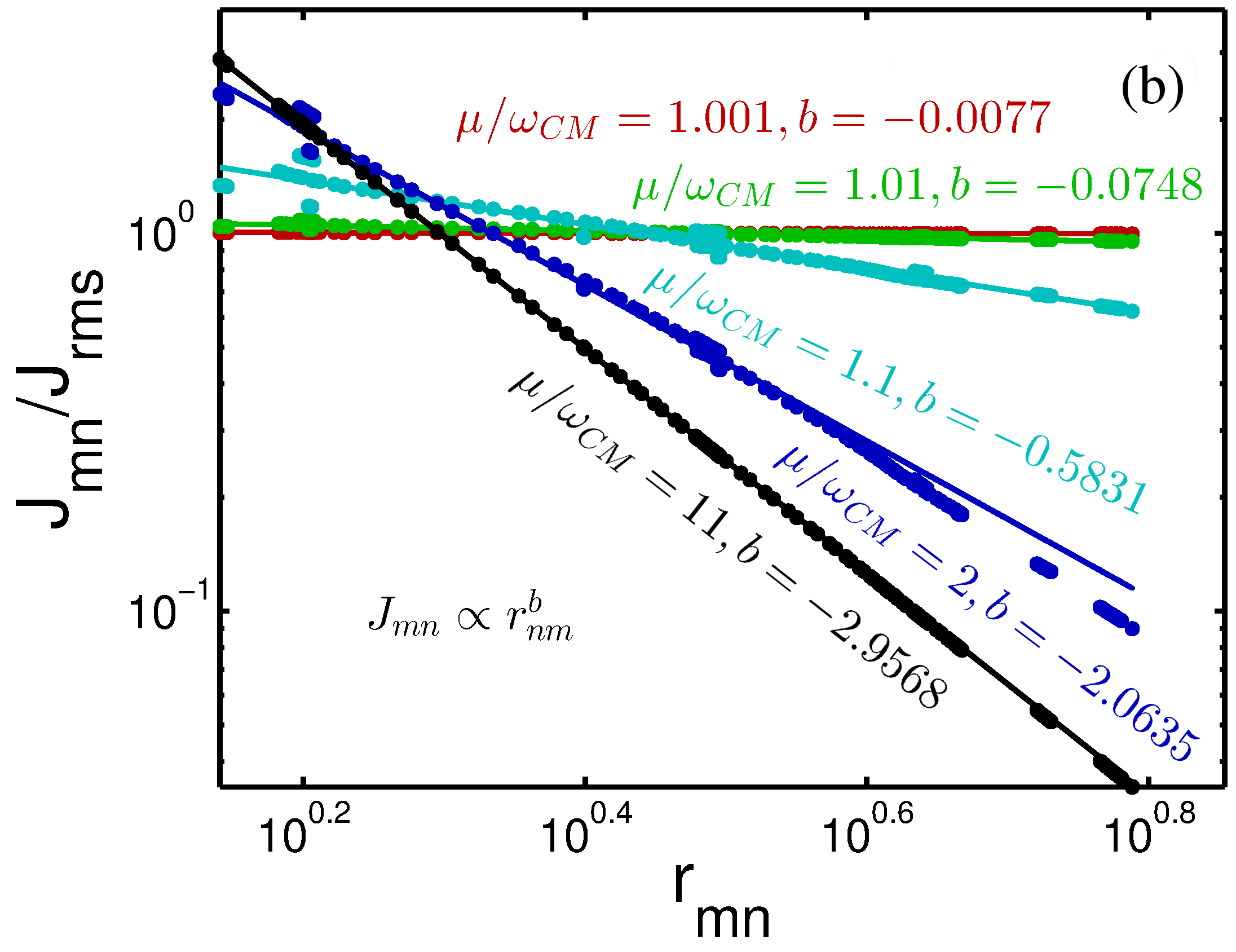}
	\caption{Spin-spin coupling versus distance for different detunings with DC voltages of $V_r=94.9$V and $V_t=V_b=100$V and different number of ions (different panels; a is $N=10$ and b is $N=20$) on a log-log plot. The colored lines are the fit to a power-law behavior $J_{mn} \propto r_{mn}^{-b}$. The detunings are $\mu=1.001$, $1.01$, $1.1$, $2$, and $11$ to the blue of $\omega_{CM}$ in units of $\omega_{CM}$.}
	\label{fig:JijV100}
\end{figure}

It is expected that if one detunes, $\mu$, to be larger than the center-of-mass mode frequency, then one can generate long-range
spin-spin couplings that vary as a power law from 0 to 3, as they decay with distance. Hence,
we fit the spin-spin couplings to a power law in distance as a function of detuning in Fig.~\ref{fig:JijV100} for $N=10$ and $20$. Note that our system is still rather small, so there are likely to be finite size effects that modify the simple power law behavior.

\subsection{Quantum motional effects}

The trap could also be used to examine different types of quantum
motional effects of ions, similar to the recent work on the
Aharonov-Bohm effect~\cite{aharonov_bohm}. In order to examine such
effects, one would need to cool the system to nearly the ground
state. This can be accomplished by including Raman side-band cooling
after Doppler cooling the system for all modes except the soft
rotational mode, at least when the potential is large enough that the
mode frequencies are sizeable.  To cool the rotational mode, one would
need to add a perturbation to the system that lifts the phonon mode
frequency, side-band cool it, and then adiabatically reduce the
frequency by removing the perturbation. This procedure will cool off
that phonon mode, which can yield quite small quanta in
it~\cite{aharonov_bohm}. Once the system has been prepared in this
state, then quantum tunneling effects, or coherent motional effects
could be studied in the trap for a range of different ion
configurations.  It might also be interesting to extend these types of
studies to cases where the ions no longer lie completely in one plane,
but have deformed into a full three-dimensional structure (as long as the larger micromotion does not cause problems).  Finally,
many of these ideas would need to be used if one tried to examine time
crystals, especially the cooling of the rotational mode to be able to
see quantum effects.

\section{Conclusion}

In this work, we have studied 2D ion crystals in an oblate Paul trap
for use in quantum simulations. With this system, one can trap a
modest number of ions in 2D planar structures that are likely to be
highly frustrated without needing a Penning trap, providing a
controlled way to study the onset of frustration effects in quantum
simulations.  We calculated the equilibrium positions and the phonon
frequencies for the proposed oblate Paul trap over its stable region. The
equilibrium positions with $N\leq5$ form a single ring configuration
and could potentially be used to study periodic boundary conditions and
the Aharonov-Bohm effect when $N=4$ or $5$ (and possibly time
crystals). Once $N > 5$, the equilibrium configurations have multiple
rings that are nearly formed from triangular lattices. One can
generate an effective Ising Hamiltonian by driving axial modes with a
spin-dependent optical dipole force. When detuning is to the blue of
the axial center-of-mass mode, the spin-spin coupling, $J_{mn}$, has an
approximate power law that is within the expected range of $0$ to $3$.
In the future, as this trap is tested and performs simulations of
spin models with ions, the work presented here will be critical to
determining the parameters of the Hamiltonian and for selecting the
appropriate configurations to use in the different simulations.
 

\begin{backmatter}

\section*{Competing interests}
  The authors declare that they have no competing interests.

\section*{Author's contributions}
The idea from this work came from Wes Campbell and Jim Freericks. Danilo Dadic and Wes Campbell developed the electric field potentials for the pseudopotential description of the trap as well as designing the trap parameters. Bryce Yoshimura, Marybeth Stork and Jim Freericks performed all of the theoretical calculations. The paper manuscript was first drafted by Bryce Yoshimura and then all authors contributed to revisions.

\section*{Acknowledgements}
We thank Dr. Philippe Bado, Dr. Mark Dugan and Dr. Christopher Schenck of Translume (Ann Arbor, MI) for valuable discussions. B. Y. acknowledges the Achievement Rewards for College Scientists Foundation for supporting this work. M. S. acknowledges the National Science Foundation under grant number DMR-1004268 for support. J. K. F. and B. Y. acknowledge the National Science Foundation under grant number PHY-1314295 for support.
D.D. and W.C.C. acknowledge support from the U.S. Air Force Office of Scientific Research Young Investigator Program under grant number FA9550-13-1-0167 and support from the AFOSR STTR Program.  J. K. F. also acknowledges support from the McDevitt bequest at Georgetown University.

\bibliographystyle{bmc-mathphys} 
\bibliography{2DPaultrap}      

\end{backmatter}
	\end{document}